\documentclass[
reprint,
superscriptaddress,
 amsmath,amssymb,
 aps,
]{revtex4-2}

\newcommand{\STO}{\text{SrTiO$_3$}}
\newcommand{\sto}{\text{SrTiO$_3$}}

\newcommand{\bzo}{\text{BaZrO$_3$}}
\newcommand{\eps}{\varepsilon}
\newcommand{\e}[1]{\ensuremath{\text{e}^{#1}}}

\usepackage{changes}

\usepackage[utf8]{inputenc}
\usepackage[english]{babel}

\usepackage{latexsym}              
\usepackage{amsmath}            
\usepackage{amssymb}
\usepackage{mathtools}
\usepackage{bbm}
\usepackage{bm}
\usepackage{physics}
\usepackage{amsfonts}
\usepackage{printlen}
\usepackage{tabularx}

\usepackage{dcolumn}

\usepackage{graphicx}
\usepackage{parskip}
\usepackage[most]{tcolorbox}

\newlength\figureheight
\newlength\figurewidth
\usepackage{listings}
\usepackage{color}
\usepackage{threeparttable}
\usepackage{graphicx}

\begin{document}

\title{Hard and soft materials: Putting consistent
van der Waals density functionals to work}

\author{Carl M. Frostenson$^*$}%
\affiliation{Department of Microtechnology and Nanoscience - MC2, Chalmers University of Technology, SE-41296 Gothenburg, Sweden}
\email{Equal contribution}
\author{Erik Jedvik Granhed$^*$}%
\affiliation{Department of Materials Science and Engineering, KTH Royal Institute of Technology, SE-10044 Stockholm, Sweden}
\affiliation{Department of Physics, Chalmers University of Technology, SE-41296 Gothenburg, Sweden}
\email{Equal contribution}
\author{Vivekanand Shukla}%
\affiliation{Department of Microtechnology and Nanoscience - MC2, Chalmers University of Technology, SE-41296 Gothenburg, Sweden}
\author{P{\"a}r A. T. Olsson}%
\affiliation{Materials Science and  Applied Mathematics,  Malmö University, SE-205  06 Malm\"o, Sweden}
\affiliation{Division of Mechanics,  Lund University, Box 118,  SE-221 00 Lund, Sweden}
\author{Elsebeth Schr{\"o}der}
\affiliation{Department of Microtechnology and Nanoscience - MC2, Chalmers University of Technology, SE-41296 Gothenburg, Sweden}
\author{Per Hyldgaard}
\affiliation{Department of Microtechnology and Nanoscience - MC2, Chalmers University of Technology, SE-41296 Gothenburg, Sweden}
\email{hyldgaar@chalmers.se}

\date{\today}

\begin{abstract}
We present the idea and illustrate potential benefits of having a tool chain of closely related regular, unscreened and screened hybrid exchange-correlation (XC) functionals, all within the consistent formulation of the van der Waals density functional (vdW-DF) method
[JPCM 32, 393001 (2020)]. Use of this chain of 
 nonempirical  XC functionals allows us to map
 when the inclusion of truly
 nonlocal exchange and of truly nonlocal correlation is important.
 Here we begin the mapping by addressing 
 hard and soft material challenges: magnetic elements, perovskites, 
and biomolecular problems.
We also predict the structure and polarization for a  ferroelectric
polymer. To facilitate this work and future broader explorations, we furthermore present
a stress formulation for spin vdW-DF and illustrate 
use of a simple stability-modeling scheme
to assert when the prediction 
of a soft mode (an imaginary-frequency
vibrational mode, ubiquitous  
in perovskites and soft matter) implies a prediction of
an actual low-temperature transformation.

\end{abstract}

\maketitle

\section{Introduction}

Modern density functional theory (DFT) 
calculations
seek to describe general matter, ideally with one and the same
exchange-correlation (XC) energy functional for all materials, i.e., 
under a general-purpose hat. Truly nonlocal and 
strong correlation effects, as well as truly nonlocal
(Fock) exchange, play 
important roles in many systems, where different 
interaction components compete 
\cite{rydberg03p606,langrethjpcm2009,bjorkmannlayered2,bearcoleluscthhy14,beckeperspective,BurkePerspective,Interface_perspective}. 
Some challenges come 
from the tendency to overly delocalize orbitals in 
regular, that is, density-explicit functionals, and 
some from the need to handle strong (local) 
correlation. These problems can be ameliorated by 
inclusion of a fraction of Fock exchange in 
so-called hybrid XC functionals 
\cite{becke93,PBE0,Perdew96,Burke97,HeyScuErn2003,DFcx02017,cx0p2018} or 
by inclusion of a Hubbard 
term, in so-called DFT+U \cite{AnZaAn91}. A further 
long-standing challenge for DFT is a proper and 
balanced inclusion of van der Waals (vdW) forces 
\cite{Dion04,langrethjpcm2009,revA,revB,revC,revD,Berland_2015:van_waals,Interface_perspective,HyJiSh20}.

It is expected that, at least for now, one must retain both a regular (density-explicit) XC functional, a hybrid XC functional, as well as an option for a Hubbard-U correction in DFT calculations \cite{BurkePerspective}. However, it is also desirable to limit the personal DFT tool box to essentially two to three fixed XC choices of related origin. This is because one can then more easily compare DFT results among different types of materials and more easily gather experience to seek further development \cite{BurkePerspective,behy14,HyJiSh20}. For example, a popular choice is to use XC functionals that originate from the constraint-based formulation of the generalized gradient approximation (GGA) \cite{lape80,lameprl1981,pewa86,pebuwa96,lavo87,lavo90}, by picking PBE \cite{pebuer96} as the 
regular functional, 
PBE0 \cite{PBE0,Burke97} as an unscreened hybrid,
and HSE \cite{HeyScuErn2003,HeyScuErn2006} as the range-separated hybrid (RSH) that
also secures a screening of the long-range Fock-exchange component. This XC tool-chain works when the impact of truly nonlocal-correlation effects can be ignored.

The vdW density functional (vdW-DF) method \cite{ryluladi00,rydberg03p126402,Dion04,Dion05,thonhauser,lee10p081101,bearcoleluscthhy14,behy14,Thonhauser_2015:spin_signature,DFcx02017,HyJiSh20} has a systematic inclusion of truly nonlocal correlation effects. Moreover, it now also provides
an XC tool chain of closely related consistent vdW-DF \cite{HyJiSh20} XC functionals. That is, the method now 
has the consistent-exchange vdW-DF-cx \cite{behy14,bearcoleluscthhy14,Thonhauser_2015:spin_signature} (here abbreviated CX) as a current-conserving
density-explicit XC starting point, the 
zero-parameter vdW-DF-cx0p \cite{DFcx02017,cx0p2018} 
(here abbreviated CX0P) as an associated unscreened nonlocal-correlation hybrid,
and the new vdW-DF-ahcx \cite{AHCX21} (here abbreviated AHCX) RSH hybrid form. 
These forms are deliberately kept free of adjustable parameters. 

The vdW-DF method and our consistent-vdW-DF tool chain include a balanced density-based account of van der Waals (vdW) forces, starting from the screening insight that reflects the design of semilocal functionals \cite{behy14,bearcoleluscthhy14,hybesc14,Berland_2015:van_waals,Thonhauser_2015:spin_signature,HyJiSh20,AHCX21}. It places all of the competing interactions on an equal ground-state DFT foundation, as all terms directly  reflect the variation in the ground state electron density  $n(\mathbf{r})$. This is true also for CX0P and AHCX, because they use the Kohn-Sham (KS) orbitals of the underlying density-explicit CX functional in the Fock-exchange evaluation \cite{behy14,HyJiSh20}. Moreover, the Fock-exchange mixing in CX0P \cite{cx0p2018} and (by 
extension) in AHCX \cite{AHCX21} is set from
an analysis of the coupling-constant scaling analysis of the
correlation-energy term \cite{signatures}, 
which again is completely
specified by the electron density variation \cite{cx0p2018}.

In this paper, we illustrate the general-purpose usefulness of the 
consistent-vdW-DF XC tool chain (CX/CX0P/AHCX), and we begin work to 
expand their use for magnetic systems.  That is, we provide a formulation 
of stress in spin vdW-DF calculations and implement  
it in the planewave-DFT software suite \textsc{Quantum Espresso} \cite{QE,Giannozzi17}. We also illustrate an approach to discuss
stability in the presence of soft modes, i.e., vibrational modes that have an imaginary frequency when described in a quadratic approximation to the potential energy variation with local deformations. Our approach is inspired by a quantum theory of temperature variations of polarization fluctuations above the ferroelectric transition temperature \cite{Mahan2013}. We combine Landau-expansion theory \cite{Callaway_book} with inelastic resonant tunneling \cite{ChEsTs74,Luryi85,JoGr87,HyJa90,DaHeHyWi93,HyHeDaWi94} for a simple, but generic, discussion of materials characterizations in the presence of soft modes. 

Accurate determinations of spin and vibrational effects are central requirements for the usefulness of the vdW-DF method. A proper spin-vdW-DF formulation for the XC value $E_{\rm c}^{\rm nl}$ and for XC-potential components, $v_{{\rm c},s=\uparrow,\downarrow}^{\rm nl}(\mathbf{r})$ is generally needed to accurately describe the atoms, and hence bulk cohesion \cite{Gharaee2017}.  However, we also need spin in many materials directly, for example, in magnetic elements and perovskites. For such problems it is desirable
to have access to spin-vdW-DF stress results to enable consistent structural optimizations. Similarly, vibrations often directly affect and will at least fine tune material characterizations, as some of us
have explicitly demonstrated for transition-metal and perovskite thermophysical properties \cite{Gharaee2017,GraLinWah2019,grwahy20}.

A broad test, from hard to soft matter, of usefulness
of the consistent-vdW-DF tool chain is needed. 
Structure, polarization, and vibrations are 
seen as strong discriminators of DFT 
performance as they directly reflect the electronic structure variation \cite{bearcoleluscthhy14,BrownAltvPRB16,HyJiSh20,C09ferro,Jia2019}. 
The CX/CX0P/AHCX demonstration and testing goal is pursued by computing material properties using at least two parts of the tool chain (as relevant and possible).
We characterize magnetic elements' structure 
and cohesion, structure in a ferromagnetic perovskite,
as well as the elastic response, vibrations, and phase 
stability in the nonmagnetic \sto . 
The latter has a known phase transition and offers an opportunity for contrasting with 
the \bzo , which remains cubic all the way down to zero 
temperature \cite{GraLinWah2019,grwahy20}.
We furhermore study biomolecular test cases and intercalation in DNA
to document that CX is accurate for soft matter,
and proceed to predict the structure and
polarization response in the ferroelectric
polyvinyl-di-fluoride (PVDF) polymer crystals.

The paper is outlined as follows. In Section II we present
theory, including a formulation of stress calculations
in spin vdW-DF, and a discussion of modeling 
phase stability in cases where XC calculations
yield soft modes. Section III presents
a number of challenges, from hard to soft, on which
we validate the theory contribution and illustrate 
use of the consistent-vdW-DF tool chain: This section
also presents the computational details as they 
pertain to our study of these challenges. Section IV and V
present our results for hard and soft
matter problems, respectively. Finally, Section VI
contains a discussion and summary. The paper has one appendix.

\section{Theory}

The vdW-DF method is in general well set up as a materials theory tool. It is, for example, implemented
in broadly used DFT code packages such as  \textsc{Quantum Espresso} \cite{QE,Giannozzi17}, \textsc{VASP} \cite{KreFur_CMS_1996,KreFur_PRB_1996}, \textsc{WIEN2K} \cite{BLAHA1990,WIEN2K2020}, \textsc{CP2K} \cite{VANDEVONDELE2005103,CP2K2020}, as well as in \textsc{GPAW} \cite{Mortensen2005,Enkovaara2010}
and \textsc{OCTOPUS} \cite{MARQUES200360,OCTOPUS06,C5CP00351B}
through our \textsc{libvdwxc} library \cite{libvdwxc}.  The code packages come with a full set of vdW-DF versions and variants.  

In some code packages the spin effects on energies, forces and stress are approximated by setting the nonlocal correlation terms without attention to spin impact on the underlying plasmon-dispersion model. This is not so in our full \textsc{Quantum Espresso} implementation of spin vdW-DF (which includes spin versions of CX, CX0P, and AHCX),
the implementation thus permits users to check if there are relevant spin 
vdW-DF effects to consider in their system of interest, for example, in the description
of bulk cohesion \cite{Gharaee2017}. 
However, to fully benefit from this \textsc{Quantum Espresso}
status, we need to enable variable-cell calculations by providing also a 
stress description for spin vdW-DF.

\subsection{Spin vdW-DF calculations}

The vdW-DF method is a systematic approach to design XC functionals that capture truly nonlocal correlation effects. Pure vdW interactions (produced by electrodynamic 
coupling of electron-hole pairs \cite{HyJiSh20}) are examples of nonlocal-correlation effects. Another example is the screening (by itinerant valence electrons \cite{HyJiSh20}) that shifts orbital energies as, for example, captured in a cumulant expansion. In the vdW-DF design we note 
that both are reflected in an electrodynamics reformulation of the XC functional. This allows us to treat all XC effects on the same footing in the electron-gas tradition. 

In practice, we use a GGA-type functional
$E_{\rm xc}^{\rm in}$ to define an effective 
(nonlocal) model of the frequency-dependence of the electron-gas susceptibility $\alpha(\omega)$. For reasons
discussed elsewhere \cite{thonhauser,hybesc14,Berland_2015:van_waals}, we limit this input to LDA plus a simple approximation for gradient-corrected exchange.
We formally express the internal semilocal functional $E_{\rm xc}^{\rm in}$
as the trace
of a plasmon propagator $S_{\rm xc}(\bm{r},\bm{r'},\omega)$,
\begin{equation}
    E_{\rm xc}^{\rm in} = \int_0^{\infty} \frac{du}{2\pi} \hbox{\rm Tr}\{
    S_{\rm xc}(\omega=iu)\} -
    E_{\rm self}\, .
    \label{eq:xcIn}
\end{equation}
The trace is here taken over the spatial coordinates of $S_{\rm xc}$. The term $E_{\rm self}$ denotes an
infinite self-energy that removes the formal divergence.

The key point is that the model plasmon propagator $S_{\rm xc}$ also defines an effective GGA-level model dielectric  function $\epsilon(iu) = \exp(S_{\rm xc}(iu))$ and a corresponding model susceptibility
$\alpha(iu)=(\epsilon(iu)-1)/4\pi$. Moreover, by enforcing current conservation, the dielectrics modeling also defines the full vdW-DF specification of the XC functional, 
\begin{equation}
    E_{\rm xc}^{\rm DF} = \int_0^{\infty} \frac{du}{2\pi} \hbox{\rm Tr}\{\ln(\nabla \epsilon_{\rm xc}(iu)\cdot \nabla G)\} -
    E_{\rm self}\, ,
    \label{eq:xcDF}
\end{equation}
where $G$ denotes the Coulomb Green function.
By expanding Eq.\ (\ref{eq:xcDF}) to first order in
$S_{\rm xc}$, one formally recoups the internal
GGA-type functional Eq.\ (\ref{eq:xcIn}).
By further expanding to second order, we obtain the vdW-DF
determination of corresponding nonlocal-correlation effects,
\begin{equation}
    E_{\rm c}^{\rm nl,sp} = \int_0^{\infty} \frac{du}{4\pi} \hbox{\rm Tr}\{S_{\rm xc}^2
    - (\nabla S_{\rm xc}\cdot \nabla G)^2\} \, .
    \label{eq:cnl}
\end{equation}
As indicated by superscript `sp', the
nonlocal-correlation term depends on the 
spatial variation in the spin polarization
$\eta(\bm{r})=[n_{\uparrow}(\bm{r})-n_{\downarrow}(\bm{r})]/[n_{\uparrow}(\bm{r})+n_{\downarrow}(\bm{r})]$.

Functionals of the vdW-DF family are generally expressed as
\begin{equation}
    E_{\rm xc}^{\rm vdW-DF\#} =
    E_{\rm xc}^{0}+E_{\rm c}^{\rm nl,sp} \, , 
    \label{eq:DFapprox}
\end{equation}
where $E_{\rm xc}^0 = E_{\rm xc}^{\rm in}+\delta E_{\rm x}^0$ contains nothing but LDA and the gradient-corrected exchange. The $E_{\rm c}^{\rm nl,sp}$ is a correlation term. The Lindhard-Dyson screening logic formally mandates that the cross-over exchange term $\delta E_{\rm x}^0$ must vanish, thus setting the balance between
exchange and correlation. There are, however,
practical limitations that prevent going directly for such fully consistent implementations directly.
In the consistent-exchange vdW-DF-cx version we have chosen $\delta E_{\rm x}^0$ so that the nonzero 
cross-over term does not 
affect binding energies, in typical 
bulk and in typical molecular-interaction cases,
as discussed separately, 
Refs.\  \onlinecite{behy14,bearcoleluscthhy14,HyJiSh20,Ageo20}.

For actual evaluations, we use a two-pole approximation for the shape of the plasmon-pole propagator $S_{\rm xc}$. This plasmon-pole description, and hence the resulting vdW-DF version, is effectively set by the choice of the semilocal internal functional $E_{\rm xc}$ via Eq.\ (\ref{eq:xcIn}). This leads to the computationally
efficient nonlocal-correlation determination
\begin{equation}
    E_{\rm c}^{\rm nl} = \frac{1}{2}\int_{\bm{r}}
    \int_{\bm{r'}} n(\bm{r}) n(\bm{r'})
    \Phi(D; q_0(\bm{r}); q_0(\bm{r'})) \, ,
    \label{eq:cnleval}
\end{equation}
where $D=|\bm{r}-\bm{r'}|$. It is given by a universal kernel form $\Phi$, as discussed in Ref.\ \onlinecite{Chapter2017}. In Eq.\ (\ref{eq:cnleval}), the values of $q_0(\bm{r})$ and  $q_0(\bm{r'})$ characterize the model plasmon dispersion. 

The above-summarized vdW-DF framework leaves no ambiguity about 
how to incorporate spin-polarization effects
in $E_{\rm c}^{\rm nl,sp}$. 
Spin enters via the exchange and via the LDA-correlation
parts of $E_{\rm xc}^{\rm in}$, as given by the
spin-scaling description and by the now-standard PW92 
formulation of LDA \cite{pewa92}. This, in turn, determines the form of $S_{\rm xc}$ and ultimately
Eq.\ (\ref{eq:cnleval}). Specifically, 
the values of $q_0(\bm{r})$ and $q_0(\bm{r'})$ are set from the local energy density of the 
internal semi-local functional to reflect  
the density and spin impact on the underlying
screening description.  Of course, it is imperative to also include spin in the term $E_{\rm xc}^0$.

\subsection{The nonlocal-correlation stress tensor}

For non-spin-polarized problems, there has since long existed a formulation of stress in vdW-DF calculation \cite{sabatini12p424209}. This 
allows effective KS structure optimizations as implemented in \textsc{Quantum Espresso}.  We now present a spin vdW-DF extension of stress to enhance the KS-structure search part. It is based on the ideas of Nielsen and Martin \cite{Nielsen1985} and we first summarize the non-spin vdW-DF stress calculations, as derived
by Sabatini and co-workers \cite{sabatini12p424209}. We consider the impact of unit-cell and coordinate scaling, for example, as expressed in cartesian coordinates for a position vector $r_{\alpha} \to \tilde{r}_{\alpha}= \sum_\beta (\delta_{\alpha,\beta}+\varepsilon_{\alpha,\beta}) r_\beta$, where $\varepsilon_{\alpha,\beta}$ is the strain tensor. This scaling affects the double Jacobian, the total-electron density $n(\bm{r'})$, the total-density gradient $\nabla n(\bm{r})$ and the 
coordinate-separation variable $D$ inside
$\Phi$ in Eq.\ (\ref{eq:cnleval}). Details of these different scaling effects are discussed in  Appendix A for the spin-polarized case. 

In the absence of spin polarization, Sabatini and co-workers
\cite{sabatini12p424209} derived  the nonlocal-correlation stress tensor contribution
\begin{eqnarray}
    \sigma_{c,\alpha,\beta}^{\rm nl}
    & = & \delta_{\alpha,\beta} \left[2 E_c^{\rm nl} -
    \int n(\bm{r}) v_{c}^{\rm nl}(\bm{r})\right] \nonumber \\
    & & +\frac{1}{2}\int_{\bm{r}} \int_{\bm{r'}}
    n(\bm{r}) n(\bm{r'}) \frac{\partial \Phi}{\partial D}
    \, C_{\alpha,\beta}(\bm{r},\bm{r'}) \nonumber \\
    & & - \int_{\bm{r}} \int_{\bm{r'}}
    n(\bm{r}) n(\bm{r'}) \frac{\partial \Phi}{\partial q_0} \, 
    G_{\alpha,\beta}(\bm{r}) \, , 
    \label{eq:Sabatini} 
\end{eqnarray}
where
\begin{equation}
    C_{\alpha,\beta}(\bm{r},\bm{r'})  =  {(r_{\alpha}-r'_{\alpha})(r_\beta-r'_\beta)}/D \, ,
    \label{eq:SabatiniD}
\end{equation}
and
\begin{equation}
    G_{\alpha,\beta}(\bm{r})  =  \frac{\partial q_0(\bm{r})}{\partial |\nabla n(\bm{r})|}
    \, \frac{(\partial n(\bm{r})/\partial r_\alpha)(\partial n(\bm{r})/\partial r_\beta)}{|\nabla n(\bm{r})|} \, . 
    \label{eq:SabatiniG}
\end{equation}
This stress component is in part given by the nonlocal-correlation contributions, 
$v_{c}^{\rm nl}(\bm{r})$ and $E_{\rm c}^{\rm nl}$, to the XC potential and XC energy.
These contributions are given
by local values of an inverse length
scale, $q_0(\bm{r})$, that determines
the local plasmon dispersion. As such, 
the contributions depend on the density gradients and hence have an indirect 
dependence on coordinate scaling, 
as summarized in Eq.\ (\ref{eq:SabatiniG}).

For actual evaluations it is important to note that the density gradient will scale 
in unit-cell and coordinates,
both since the density scales with the unit-cell size and because the formal expression 
for the spatial gradient scales with coordinate dilation even at a fixed density.  
The former effect is incorporated by the term containing $v_{c}^{\rm nl}(\bm{r})$. 
The latter effect is captured by the term in the last row of Eq.\ (\ref{eq:Sabatini}). 
Fortunately, the local variation of the inverse length scale $q_0$ is already computed 
as part of any self-consistent vdW-DF calculation. 

The proper spin vdW-DF formulation is presently implemented in \textsc{Quantum Espresso},
in \textsc{WIEN2K} and in our \textsc{libvdwxc} (and thus
available also in both \textsc{GPAW} and 
\textsc{Octopus}). For spin-carrying systems we generally work with the spin-up and spin-down density components $n_{s=\uparrow,\downarrow}(\bm{r})$ of the total electron density $n(\bm{r})=n_{\uparrow}(\bm{r})+n_{\downarrow}(\bm{r})$. 
The internal functional depends on the variation in the spin polarization $\eta(\bm{r})$
and this spin dependence impacts the plasmon 
dispersion, and ultimately  $E_{\rm c}^{\rm nl,sp}$, through the local $q_0(\bm{r})$ values.  

To also compute stresses in spin vdW-DF we must update Eq.\ (\ref{eq:Sabatini}) accordingly;
Details are discussed in Appendix A. 
We find that the second line of Eq.\ (\ref{eq:Sabatini}) 
only changes to the extent that the values of the $q_0$'s
must now be evaluated for $\eta(\bm{r})\neq 0$. For the first line, there is a small modification since separate XC potentials now act on $n_{\uparrow}(\bm{r})$ and
$n_{\downarrow}(\bm{r})$. 

Finally, for an update of the third line of Eq.\ (\ref{eq:Sabatini}), we simply track the variation of the local $q_0$ values on both spin-density gradient terms. Thus, the resulting spin-vdW-DF stress tensor expression becomes
\begin{eqnarray}
    \sigma_{c,\alpha,\beta}^{\rm nl,sp}
    & = & \delta_{\alpha,\beta} \left[2 E_c^{\rm nl} -
    \sum_{s=\uparrow,\downarrow} \int n_s(\bm{r}) v_{c,s}^{\rm nl,sp}(\bm{r})\right] \nonumber \\
    & & +\frac{1}{2}\int_{\bm{r}} \int_{\bm{r'}}
    n(\bm{r}) n(\bm{r'}) \frac{\partial \Phi}{\partial D}
    \, C_{\alpha,\beta}(\bm{r},\bm{r'}) \nonumber \\
    & & - \int_{\bm{r}} \int_{\bm{r'}}
    n(\bm{r}) n(\bm{r'}) \frac{\partial \Phi}{\partial q_0} \, 
    \sum_{s=\uparrow,\downarrow}
    G^s_{\alpha,\beta}(\bm{r}) \, ,
    \label{eq:newStress}
\end{eqnarray}
where
\begin{equation}
    G^{s=\uparrow,\downarrow}_{\alpha,\beta}(\bm{r}) = 
    \frac{\partial q_0(\bm{r})}{\partial |\nabla n_s(\bm{r})|} 
    \, \frac{(\partial n_s/\partial r_\alpha)(\partial n_s/\partial r_\beta)}{|\nabla n|} \, . 
    \label{eq:newStressG}
\end{equation}
As in the non-spin-polarized case,
the spin vdW-DF stress contribution, Eq.\ (\ref{eq:newStressG}), is conveniently given by quantities that are already computed in a self-consistent determination of the electron density variation in spin vdW-DF.

Sabatini \emph{et al.} \cite{sabatini12p424209} incorporated their non-spin  result in the \textsc{Quantum Espresso} package, and we have  implemented and released our here-described spin extension in the same package.

\subsection{Hybrids based on GGA and vdW-DF}

In total, eight different functionals were used for the calculations. In some perovskite studies we include results obtained by the local-density approximation (LDA) \cite{pewa92}
for reference. For studies with density-explicit GGA and vdW-DF, we 
generally rely on the PBE  \cite{pebuer96} and CX
\cite{behy14,bearcoleluscthhy14} versions; For polymers we also compare CX results to vdW-DF1 \cite{Dion04} and vdW-DF2 \cite{lee10p081101} results. 

For some hard and soft systems, we use and compare results obtained in PBE- and CX-associated hybrids, both unscreened and in a RSH form. 
The vdW-DF-cx0 class is an unscreened
hybrid class that is formulated in analogy with the PBE0 \cite{PBE0,Burke97,DFcx02017}. We use
the zero-parameter form CX0P \cite{cx0p2018} in which the extent of Fock exchange mixing is kept fixed at 20\%, 
following an analysis of the CX coupling-constant scaling \cite{signatures} that enters the general hybrid design logic \cite{Burke97,cx0p2018}.

The HSE functional \cite{HeyScuErn2003,HeyScuErn2006} is a RSH extension of the PBE functional. We use it with 25\% Fock exchange and a range separation that is described by a screening parameter $\mu = 0.2$~{\AA}$^{-1}$. This parameter defines an error-function weighting $\mathrm{erf}(\mu r)/r$ of the Coulomb interaction \cite{HeyScuErn2003}, thus limiting the Fock-exchange inclusion
to short separations.

Finally, we have also used the recently launched CX-based RSH, termed AHCX \cite{AHCX21},
for a study of magnetic element
structure and cohesion, as well as for some molecular benchmark studies. It resembles
the CX0P in that we keep the Fock exchange
fraction fixed at 20\% and it resembles
HSE in that we keep the screening parameter
fixed at the standard HSE value \cite{HeyScuErn2006}. The screening
makes the AHCX calculations relevant for
metallic systems \cite{AHCX21}.

\begin{figure}[htpb]
    \includegraphics[width=0.40\textwidth]{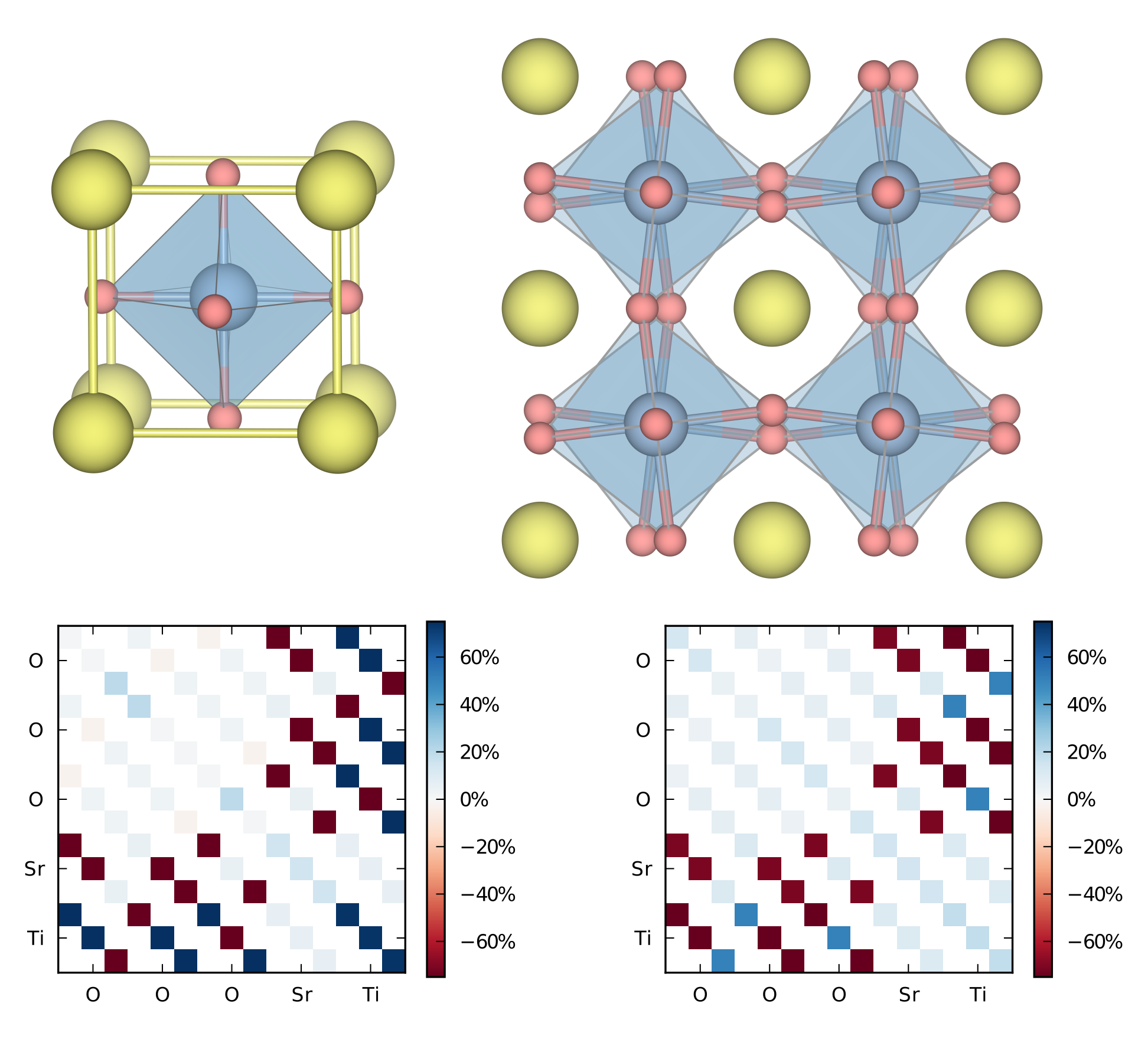}
    \caption{Top row depicts the conventional unit cell of SrTiO$_3$ (left) and a 40-atom cell (being a $2\times2\times2$ repetition)
    model that illustrates the oxygen dynamics in an important vibrational excitation at the $R$ point of the Brillouin zone, involving rigid rotations of the oxygen octahedra. This is the AFD mode, experimentally known to drive
    a phase-transition below 105~K. The bottom row shows a comparison of relative changes in the dynamical matrix (evaluated at the $\Gamma$ point) when going from PBE and to CX (left) and from CX to CX0P (right) for SrTiO$_3$. }
    \label{fig:PerovStruct}
\end{figure}

\subsection{Navigating phase transformations}

Hard and soft matter come in different
crystal forms, as well as meta-stable variants,
and there is often a need for a 
concerted theory-experiment analysis to resolve
and understand phase stability. This is true,
for example, even in simpler (compact-unit-cell
and nonmagnetic) perovskites \cite{Tien1967,Thomas1968a,VanGool1969,Saifi1970,Salje1990,Nakamura_Ferroelectrics_1992,ZhoVanRab1994,ZhoVan1995,ZhoVanRab1995,ZhoVan1996,Varnhorst1996,Vanderbilt1998,Meyer2001,Resta1993,Meyer2001,Wu2005,SalGueBou2011,TadTsu15,TadTsu19,GraLinWah2019,grwahy20}
and in both nonpolar and ferroelectric polymers \cite{Lando1966,Hasegawa1972,Resta2000,Nakamura2001,Ranjan2007,Ranjan2012,Itoh2014,Zhao2016,Dong2016,Pelizza2016,Olsson17,OlHySc18,Ruan2018,Pelizza2019}. The
structural transformations may be driven by temperature, 
electric fields, or strain; There can also be a release from
meta-stable states that may have been locked in under synthesis 
or produced by usage \cite{OlHySc18}.

First-principle calculations (that also accurately determine
stress) may help in the analysis with volume-constrained variable-cell calculations 
and by determination of the phonon spectra, as well
as calculations of magnetic, elastic, polarization, and strain
response \cite{VanKin1993,Resta1993,Togo2008,Togo2010,Tanaka2010,Pie2011,Togo2015,TadTsu15,Skelton2016,Skelton2017,OlHySc18,ErFrEr2019,FraEriErh2019,TadTsu19,Pelizza2019}. 
For a given XC functional, we can rely on the 
Born-Oppenheimer (BO) approximation to determine what we call the native structure. This approach  allows us to  
 track the dependence of, for example,  the 
\bzo\ form as a function of volume or pressure \cite{grwahy20}. We can also compute the phonon spectra at the native structure or 
at the experimental 
structure and, for example, check for soft modes
\cite{GraLinWah2019,grwahy20}. Such first principle calculations 
help us identify the nature of a potential or actual instability, for example, when the
cubic-system anti-ferrodistortive (AFD) 
mode is a potential driver for structural  transformations \cite{GraLinWah2019,grwahy20}.

Similarly, we may for polymer crystals track 
the deformation that arises with electric fields \cite{Ranjan2007}
or with applied strain \cite{OlHySc18}. Here again first-principle
calculations can identify not only the nature of modes and displacements, leading to a potential instability \cite{Olsson17,OlHySc18}, but also quantify the energy landscape 
for individual-strain or crystal transformations \cite{OlHySc18}.

However, some effective modeling beyond the first principle calculations of structure and modes is still critically needed.  
This is because the finding of an imaginary frequency simply says 
that -- for the chosen XC functional -- there is an incipient instability; Treatments of zero-temperature and temporal 
fluctuations (correcting the underlying BO 
description) are essential to assert if that XC functional 
predicts an actual phase transformation \cite{Vanderbilt1998,TadTsu15,TadTsu19} or facilitates
a polymer breakdown of long-range phase order \cite{OlHySc18}.
For the simpler, compact-structure perovskites (like \sto ), there exists both Monte-Carlo simulation frameworks \cite{ZhoVan1995,ZhoVan1996,Vanderbilt1998,Wu2005}
and a phonon Green function formulation \cite{TadTsu15,TadTsu19},
but something simpler is, in general, desirable
to limit the computational load in complex systems.

\subsection{Vibrations and stability: A simple analysis} 

Here, we illustrate use of a model analysis of quantum effects on transformations. The approach 
is generic to stability problems. Our 
model is inspired by a quantum theory of 
fluctuations above the phase-transition 
temperature \cite{Mahan2013}. Our analysis 
checks, for a given choice of 
XC functional, whether a DFT finding of 
a soft mode also implies the likely 
prediction of an actual transformation 
at $T\to 0$. Our modeling approach supplements 
the temperature-variation focus of Ref.\
\onlinecite{Mahan2013} by also taking
tunneling-induced vibrational-mode-level 
splitting into account. We focus the 
discussion on the \sto\ AFD mode, and compare DFT characterizations
obtained for a string of XC functionals.

The top right panel of Fig.\ \ref{fig:PerovStruct} shows schematics
of the \sto\ atomic configuration in its cubic, high-temperature form; The top right panel illustrates the AFD mode, with compensating 
oxygen (red spheres) rotations. This mode causes a phase transition 
out of the cubic phase at 105 K in \sto; This phase transition
does not occur in the similar \bzo\ systems \cite{GraLinWah2019,grwahy20}. The mode is located
at the $R$ position of the Brillouin zone of the simple-cubic
\sto\ form \cite{SetCurBrillioun}. It competes in \sto\ with
a $\Gamma$-point ferroelectric-instability
mode that involves shifts of the Sr atoms 
(yellow spheres), but the $R$ mode instability
is experimentally known to suppress the latter
instability \cite{ZhoVanRab1995,Vanderbilt1998,Wu2005,TadTsu19}.

The bottom row of Fig.\ \ref{fig:PerovStruct}
emphasizes a general motivation factor for testing
our CX/CX0P/AHCX tool chain on perovskites in general
and for \sto\ and \bzo\ in particular \cite{GraLinWah2019,grwahy20}.
The point is that the description of the interatomic forces 
(defining both the $\Gamma$- and $R$-mode phonons) differs 
significantly as we change between a choice of PBE-CX
and CX-CX0P XC functionals. Specifically, the bottom-right
panel lists, in color coding for all atom pairs (and the cartesian coordinates of the resulting forces, i.e., 15 by 15 entries in total) the relative  change in forces as we go from CX to CX0P. A similarly large impact of the XC functional nature was previously documented for \bzo, cf.\ Fig.\ 4 of Ref.\ \onlinecite{grwahy20}. The differences (that affect the
description of vibrations) exist even if all functionals accurately predict the cubic lattice constant.

\begin{figure}[ht]
	\includegraphics[width=0.93\columnwidth]{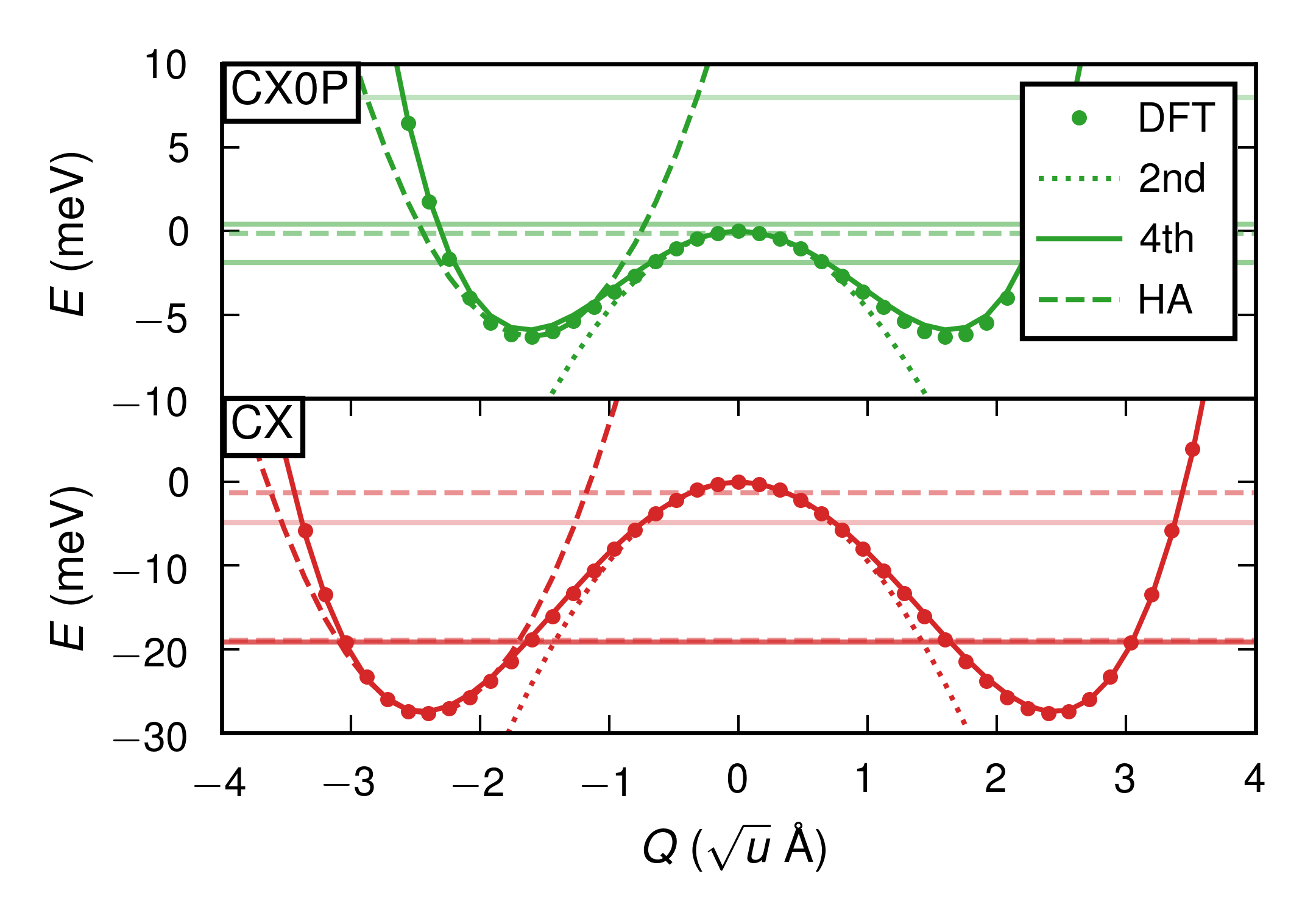}
	\caption{
	Potential energy landscape along a distortion coordinate $Q$ representing the $R_{25}$ phonon mode, as computed and modeled  with CX0P (top panel) and
	CX (bottom panel).
	The solid  curves show a 4th order fit to frozen-structure DFT results (marked with large dots) obtained for deformations 
	in the range  $Q \in [-3,3]$ Å$\sqrt{\mbox{u}}$ for a 40-atom cell. 
	The dotted and dashed parabolae show 2nd-order  fits to  the three data points (having atomic displacements of $0.01$~Å) closest to $Q=0$ and for the displaced minima, denoted $Q_0$ ($Q_0\approx -2.5$ Å$\sqrt{\mbox{u}}$ for CX): The latter is discussed as
	the harmonic approximation (HA) and it reflects the dynamics if the specific XC functional  corresponds to a prediction of an actual low-temperature transformation.
	The solid and dashed horizontal lines indicate the energy levels obtained from the numeric solution to 
	 Eq. \eqref{eq:1DSE} as described in the full deformation and in the HA potentials, respectively.
    \label{fig:double_well_potential}
	}
\end{figure}

The important step in a simple DFT-based
stability modeling is to set up a size-consistent description of the vibrational mode that may drive distortions, for example, of the oxygen dynamics in the AFD mode in \sto. The Hamiltonian for the ionic motion (in the BO approximation) in generalised canonical coordinates (denoted $Q_{s\bf{q}}$), can be written
\begin{equation}
    H = T + V = \frac{1}{2} \sum_{s\bf{q}} \dot{Q}_{s\bf{q}}\dot{Q}_{s\bf{q}}^* +  V \, ,
    \label{eq:1DSE}
\end{equation}
The band index is $s$, and $\bf{q}$ denotes 
the phonon momentum.
The set of $Q_{s\bf{q}}$'s reflect the atomic displacements $\bm{d}_{s\bf{q}}^{n}$ and 
the phonon eigenvector $\bm{v}^{n}_{s\mathbf{q}}$ 
through the relation
\begin{equation}
    d^{n,\eta}_{s\mathbf{q}} = \frac{1}{\sqrt{\mathcal{N}M_n}} \sum_{s\bf{q}}Q_{s\mathbf{q}} \bm{v}^{n,\eta}_{s\mathbf{q}} \e{i \mathbf{q} \cdot \mathbf{R}_n} \, ,
    \label{eq:uQ-relation}
\end{equation}
where $\eta$ denotes the cartesian coordinate.
Here $\mathcal{N}$ is the number of assumed Born-von Karman repetitions, while $\mathbf{R}_n $ and $M_n$ denote the position and mass of the $n$'th atom in the unit cell.
For studying the stability of the AFD mode in the cubic structure only the momentum at the $R$ point is needed. Also, as   
illustrated in Fig.\ \ref{fig:PerovStruct}, 
the AFD modes exclusively express oxygen rotations, leaving only 
oxygen-related terms in (\ref{eq:uQ-relation}). Thus we can set $M_n=M\equiv 15.999$ atomic mass units (which we denote u).

For a non-interacting single-$R$-phonon mode, the potential energy variation can be approximated to the fourth order in the AFD displacement $d$, or equivalently in the canonical coordinate in the specific mode, $Q$ \cite{Callaway_book}. The
resulting Landau expansion is
 \begin{equation}
	T+V = \tfrac{1}{2} \dot{Q}^2 + \tfrac{1}{2} \bar{\kappa} Q^2 + \tfrac{1}{4\mathcal{N}} \bar{\alpha} Q^4 \, .
	\label{eq:Landau-expansion}
\end{equation}
Here, $\bar{\kappa}$ and $\bar{\alpha}$ are effective material-specific constants of the Landau expansion for the AFD mode, reflecting both the effective interatomic couplings and
inertia of the atoms involved in the
dynamics \cite{Callaway_book}. The overbar is
used to distinguish these parameters from
those characterizing a (related) expansion 
the potential energy surface (PES) expressed
in terms of AFD-type distortions 
(here denoted $d$)
\cite{GraLinWah2019,grwahy20}, rather than phonon-mode coordinate $Q$.

The top (bottom) panel of Fig.\ \ref{fig:double_well_potential} shows
CX0P (CX) results for the total energy variation that occurs in \sto\ when we
track the AFD-type distortion (upper right panel of Fig.\ \ref{fig:PerovStruct}) while keeping the unit-cell lattice constants
fixed at the optimal CX0P (CX) value. The dots show
DFT results for the associated deformation
energy as obtained for a 40 atom cell. 
The parameters $\bar{\kappa}$ and 
$\bar{\alpha}$ characterizing a fourth-order fit (solid curve) to the
computed potential variations
are presented in the result section IV.A. (in Table \ref{tab:main_table}). 

For our parameter fittings, we
note that Fig.\ \ref{fig:double_well_potential}
shows the total-energy variation
(in a single cell of 40 atoms)
as a function of the scaled 
AFD mode coordinate $Q$. The AFD mode in \STO\ exclusively involves 16 identical oxygen and the same values, denoted $d$ and $v$, define the magnitudes of the relevant oxygen displacements $d$ and the relevant eigenvector components $v$. Normalization mandates that $v^2=1/16$ and, in this simplified description,
Eq.\ (\ref{eq:uQ-relation}) formally reduces to $Q=\sqrt{M\mathcal{N}}d/v$. 
We use $\mathcal{N}=1$ for defining the coordinate scaling $Q$, as we fit 
$\bar{\kappa}$ and $\bar{\alpha}$ to the 40-atom PES shown in Fig.\ \ref{fig:double_well_potential}.
The relation among quadratic coefficients ($\kappa$ and $\bar{\kappa}$) in the related PES
expansions, $V \approx \kappa d^2/2= \bar{\kappa} Q^2/2$, is $\bar{\kappa}=\kappa/16M$.  

The potential energy in Eq.\ (\ref{eq:Landau-expansion}) scales properly with respect to the chosen size of the supercell or 
Born-von Karman representation, i.e.,
with $\mathcal{N}$. 
A doubling of the simulation cell in every direction ($\mathcal{N} \to 8\mathcal{N}$) yields the coordinate
rescaling, $Q \to \sqrt{8}Q$, and
the potential-energy rescaling
\begin{equation}
V = \frac{\bar{\kappa}}{2}Q^2 + \frac{\bar{\alpha}}{4\mathcal{N}}Q^4 
\to 8\frac{\bar{\kappa}}{2}Q^2 + \frac{\bar{\alpha}}{4 \cdot 8}64 \cdot Q^4 = 8V\, .
\end{equation}
Size consistency also holds for the kinetic-energy part and the set of solution vibrational frequencies, namely the phonon-type eigenlevels $\omega_{i}$. These frequencies characterize the model dynamics Eq.\ (\ref{eq:Landau-expansion}) and
they remain unchanged regardless of the choice of 
supercell size used in our DFT characterizations. 

To provide a simple assessment of which functionals are consistent with an actual low-temperature
transition (for a given material problem) we can compute (or measure) the potential-energy
surface for deformations, for example, Fig.\ \ref{fig:double_well_potential}
(or Refs.\  \onlinecite{GraLinWah2019,grwahy20}), to thus
set the Landau description, Eq.\ (\ref{eq:Landau-expansion}).
The second-order expansion parameter $\bar{\kappa}$ can sometimes 
(for example, in \bzo) give a leading-order estimate, 
$\omega_{i=1}\approx \sqrt{\bar{\kappa}}$, but a description of the
fourth-order expansion term $\bar{\alpha}$
is essential when $\bar{\kappa} < 0$ (as we shall
document holds for \sto\ in all here-investigated functional cases). We then have a potential or incipient instability and further analysis is required. 

The panels of Fig.\ \ref{fig:double_well_potential}
also show the AFD-type eigenvalue modes that result from solving for 
modal eigenlevels $\omega_i$ in the \sto\ Landau modeling based on CX0P
and CX calculations. The dashed horizontal lines describe the model eigenstate under the assumption that
the system is actually deformed
and the AFD dynamics occurs
as trapped in one of the two 
displaced harmonic oscillators described by
$Q = \pm Q_0$ (as illustrated by a dashed parabola). The pair of solid horizontal lines 
-- in each panel -- show the eigenlevels for the AFD-modal dynamics as described under the assumption that there is no relevant dephasing 
of the modal double-well dynamics \cite{Luryi85,JoGr87,DaHeHyWi93}.
That is, this double-well eigenvalue description of the vibration is provided 
under the condition that the mode retains coherence and thus exists
on both sides of the central barrier at $Q\sim 0$.

The computed value of the coefficient $\bar{\kappa}$ in the Landau expansion identifies the presence of an incipient instability for 
cubic \sto\ (the configuration described by $Q=0$) in all of the LDA, PBE, CX, HSE, and CX0p functionals. However, we must also consider the fourth-order term. The double-well shapes produce instead phonon-mode eigenvalues $\omega_i$ with a splitting, denoted 2$\Delta$. The splitting is inversely related to the depth of the double well: It is very small for LDA, fairly large for CX0P and large for HSE. However, a negative $\bar{\kappa}$ value does not necessarily imply a prediction of an  actual deformation.

To set up a simple stability criterion, we consider the \sto\ (and general incipient-instability) case as an inelastic tunneling problem
\cite{Luryi85,JoGr87,HyJa90,DaHeHyWi93,HyHeDaWi94}. We use the modal 
splitting to define a characteristic dwell or tunneling time:
\begin{equation}
\tau_R = \hbar / \Delta E \, .
\label{eq:modaltimeDef}
\end{equation}
We compare with
an assumed inelastic-scattering or dephasing time $\tau_{\rm scat}$.  A crude indicator 
for having a $T\to 0$ phase transition is
\begin{equation}
    \tau_R \gg \tau_{\rm scat} \, .
    \label{eq:stabilityCond}
\end{equation}
In essence, this is a competition 
between the phase-coherence life time and the tunneling dynamics of the  mode that could possibly drive a transformation.  

We motivate the stability condition Eq.\ (\ref{eq:stabilityCond}) as follows: 
Tunneling, as well as thermally activated fluctuations, will connect the dynamics in 
both wells.  The tunneling might be so slow that dephasing scattering occurs, preventing the mode from maintaining the coherence that exists in an isolated  quantum-mechanical double-well problem. 
In that case the inter-well dynamics is instead
exclusively caused by thermal activation 
and there will (at $T=0$) be a lock in into one of the wells. 
We interpret this lock-in as corresponding
to an actual low-temperature transformation
as $T \to 0$, and such transformation occurs in an LDA description of \sto\ and \bzo. It will also happen in 
CX, but it is not a result that emerges in
our CX0P study of \sto\ (as detailed below).
The CX0P does have a soft mode at $R$ and hence
it predicts an incipient transition. However, 
the characteristic tunneling time (predicted in 
Section IV.A) suggests that dephasing is too slow to inhibit 
tunneling in the $T\to 0$ limit. In 
a CX0P-based description, we expect that tunneling prevents an actual low-temperature phase transformation. 

The overall idea is perhaps best illustrated 
by an analogy to first-principle-theory-based analysis of addimer diffusion on 
metals, a problem that also provide a measured estimate for $\tau_{\rm scat}\sim 1$ ps \cite{ReMeRiHy03}. In the adatom/addimer-diffusion problems, for example, explored in Refs.\   \onlinecite{Stranick94p99,Stranick95p41,BoHyWaLu98,HyPe2000,ReMoRi2000,berland09p155431,Han12p19}, the dynamics eventually rolls over to a tunneling regime \cite{BoHyWaLu98}. A scanning-tunneling microscopy study documents that the addimer dynamics will never freeze out \cite{ReMeRiHy03}.
The roll over to quantum-tunneling transport occurs below $T_{\rm cross}$ = 5 K, corresponding to
$k_BT_{\rm cross}\approx 1$ meV. This energy scale
sets the time scale for $\tau_{\rm scat}
=\hbar/k_B T_{\rm cross} \approx 0.7$ ps. The
low-temperature dynamics of Cu(111) addimers 
can not be seen as thermally activated hopping 
between sites, i.e., the type of hopping (between 
dynamics in local wells) that eventually gets stuck 
in one or the other configuration. Rather, the 
dynamics maintains phase coherence and no actual lock in or trapping occurs \cite{BoHyWaLu98,ReMeRiHy03}.

Our stability estimate Eq.\ (\ref{eq:stabilityCond}) explains CX0P and CX stability differences, Section IV.A. It predicts that for CX0P-based (for CX-based) modeling of \sto, a tunneling regime will emerge (will not be relevant) at $T\to 0$. This means that the computed
imaginary frequency value of the AFD mode is unlikely
(likely) to drive an actual low-temperature transition in
a CX0P (CX) characterization of \STO .

\section{Material Challenges and Computational Details}

Our results are obtained using 
both the (stress-updated) \textsc{Quantum Espresso} 
DFT-code package \cite{QE,Giannozzi17} and
\textsc{vasp} with the setup of projector augmented wave potentials \cite{KreFur_CMS_1996,KreFur_PRB_1996}; Details are described as relevant in the subsections below. Ferroelectricity is a theme in our 
demonstrations and we use the modern (Berry-phase) theory of 
polarization \cite{VanKin1993,Resta1993}.
We rely on both the \textsc{Quantum Espresso} and \textsc{vasp} implementations \cite{KinVan1993,BarRes1986,NunGon2001,SouIniVan2002,GajHumKre2006}, 
reflecting code choices used for the specific hard and soft-matter problem.

\subsection{SrTiO$_3$ properties}

The problem of separating predictions of incipient and actual phase 
transitions motivates a detailed comparison of the tool-chain 
performance for \sto, as an illustration. Again, Fig.\ \ref{fig:PerovStruct} shows the high-temperature cubic form,
identifies the nature of the AFD distortion (as an 
$R$-mode instability, whether incipient or actual)
and highlights that the CX-based XC tool chain 
yields significantly different descriptions of 
interatomic forces than what arises in PBE.

Our study also includes calculations of phonon 
frequencies and elastic coefficients and can in part  be 
seen as an extension of a previous 
\bzo\ characterization \cite{GraLinWah2019,grwahy20}, by here 
looking at properties at the experimentally-observed 
lattice constant. However, since the \sto\ is known 
to have an actual phase transition at 105 K, we cannot track
the thermal impact on structure. Rather, we focus on discussing
which functionals, if any, are consistent with this
observation of a known \sto\ instability, driven by the AFD mode.

Accordingly, we study the energy variations of a AFD-type distortion, as well as elastic-coefficients and high-frequency dielectric 
constants in \bzo\ and \sto, as illustrated in Fig.\ \ref{fig:PerovStruct}. To this end, we use the PAW method and the \textsc{vasp} \cite{KreFur_CMS_1996,KreFur_PRB_1996} software
as well as \textsc{phonopy} \cite{Togo2015} to determine the vibrational
modes and frequencies in the cubic form, noting a need for 
additional modeling when the direct computation yields
imaginary frequencies.

Our \bzo\ and \sto\  studies are converged with respect to the wavefunction energy cut off. We have previously shown \cite{grwahy20} that the AFD mode is very sensitive to the oxygen PAW potential and the energy cut-off, and thus we use the hard
setup in \textsc{vasp}. For \bzo\ we use an energy cut-off of 1200~eV for LDA and GGA and 1600~eV for truly nonlocal functionals, i.e., HSE, CX, and CX0P.
For \sto\ energy cut-offs at 1600~eV are used for all functionals.
Convergence turns out to be less sensitive to the $k$-point sampling and a $6\times6\times6$ 
Monkhorst-Pack \cite{MonPac1976} grid was deemed sufficient for the hybrid functionals, while $8\times8\times8$ was used for non-hybrid studies.
The $R_{25}$ and $\Gamma_{15}$ frequencies were computed using the frozen phonon method with the default displacement of 0.01 Å in a 
40 atom cell (being a $2\times2\times2$ repetition of the basic cell), and postprocessed in \textsc{phonopy} \cite{Togo2015}.

The one-dimensional PES were computed by mapping out the energy landscape along the $R_{25}$ phonon mode, i.e., in a Glazer angle
\cite{Glazer1972,Glazer1975} rotation around the $z$-axis, in steps of approximately 0.14 degrees.
The energy levels were then obtained by solving the one-particle Schr{\"o}dinger equation numerically in the potential given by the PES through the finite difference method. 

\begin{figure}[htbp]
    \includegraphics[width=0.4\textwidth]{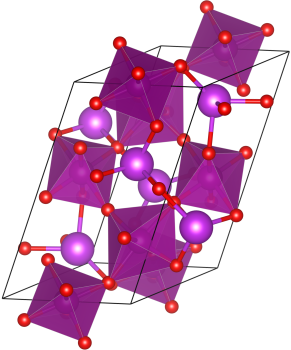} \caption{ \label{fig:structs} 
Primitive-cell representation of unit cell
and atomic configuration in the magnetic perovskite BiMnO$_3$. The unit cell has a fairly large distortion of the cubic form with the 
basis plane (described by $a$ and $b$ lattice vectors) carrying almost all of the ferroelectric response. In our schematics of the atomic configuration we use red spheres to represent
the O atoms and magenta spheres to represent
the Mn atoms that carry ferromagnetic ordering.
We also show the oxygen-octahedrals that can be seen as encapsulating the Bi atoms.}
\end{figure}

\subsection{Magnetic elements and perovskite} 

We can in general study bulk structure and compressibility properties of the magnetic Ni and Fe elements by mapping the bulk cohesive energy 
\begin{equation}
E_{\rm coh}(a) = E_{\rm bulk}(a) - E_{\rm atom}
\label{eq:cohbulk}
\end{equation} 
as a function of the assumed lattice parameter $a$. To that end we provide and compare a set of PBE, CX, and AHCX calculations in our spin-stress updated version of \textsc{Quantum Espresso}. We use optimized norm-conserving Vanderbilt~\cite{ONCV} (ONCV) pseudopotentials (PPs) in the SG15-release~\cite{sg15}. For the AHCX Fe and Ni problems, we used a  plane-wave cutoff at $200$ Ry and a $10 \times 10 \times 10$ Monkhorst-Pack~\cite{MonPac1976} $k$-point sampling,
keeping contributions from all $k$-point differences in the Fock-exchange evaluation.
We fit the results for the energy-versus-lattice constant variation to a fourth-order expansion \cite{EleniSchroder}, thus extracting the optimal lattice constant $a_{0,{\rm fit}}$, cohesive energy $E_{\rm coh}(a_{0,{\rm fit}})$, and bulk modulus $B_0$.

Variable-cell calculations involving a fraction of Fock exchange (as in hybrids) are flagged as incompletely tested in the 
 \textsc{Quantum Espresso} version that we used for the 
 AHCX launch \cite{AHCX21}; It is a question outside our focus on stress from nonlocal correlation and we limit variable-cell structure determinations to CX. In principle, such calculations give us different lattice constants, denoted $a_{0,{\rm stress}}$.
We compare $a_{0,{\rm stress}}$ and $a_{0,{\rm fit}}$ for Ni and Fe to test and validate the spin-vdW-DF stress result and implementation.

Figure \ref{fig:structs} shows schematics of the BiMnO$_3$ 
unit cell and atom configuration: O atoms (red) traps Bi atoms in 
octahedral cages in a distorted ordering, while the Mn atoms 
(magenta) carry the ferromagnetic ordering. The BiMnO$_3$ 
perovskite has a fairly large structural deformation and thus 
represents another good system for testing the new spin-vdW-DF stress implementation as well as the CX accuracy. For this problem we determined 
the structure in \textsc{Quantum Espresso} variable-cell calculations using a plane-wave cutoff of $160$ Ry and a  $10 \times 10 \times 6$ Monkhorst-Pack grid \cite{MonPac1976} to sample the Brillouin zone.

\begin{figure}
\includegraphics[width=0.47\textwidth]{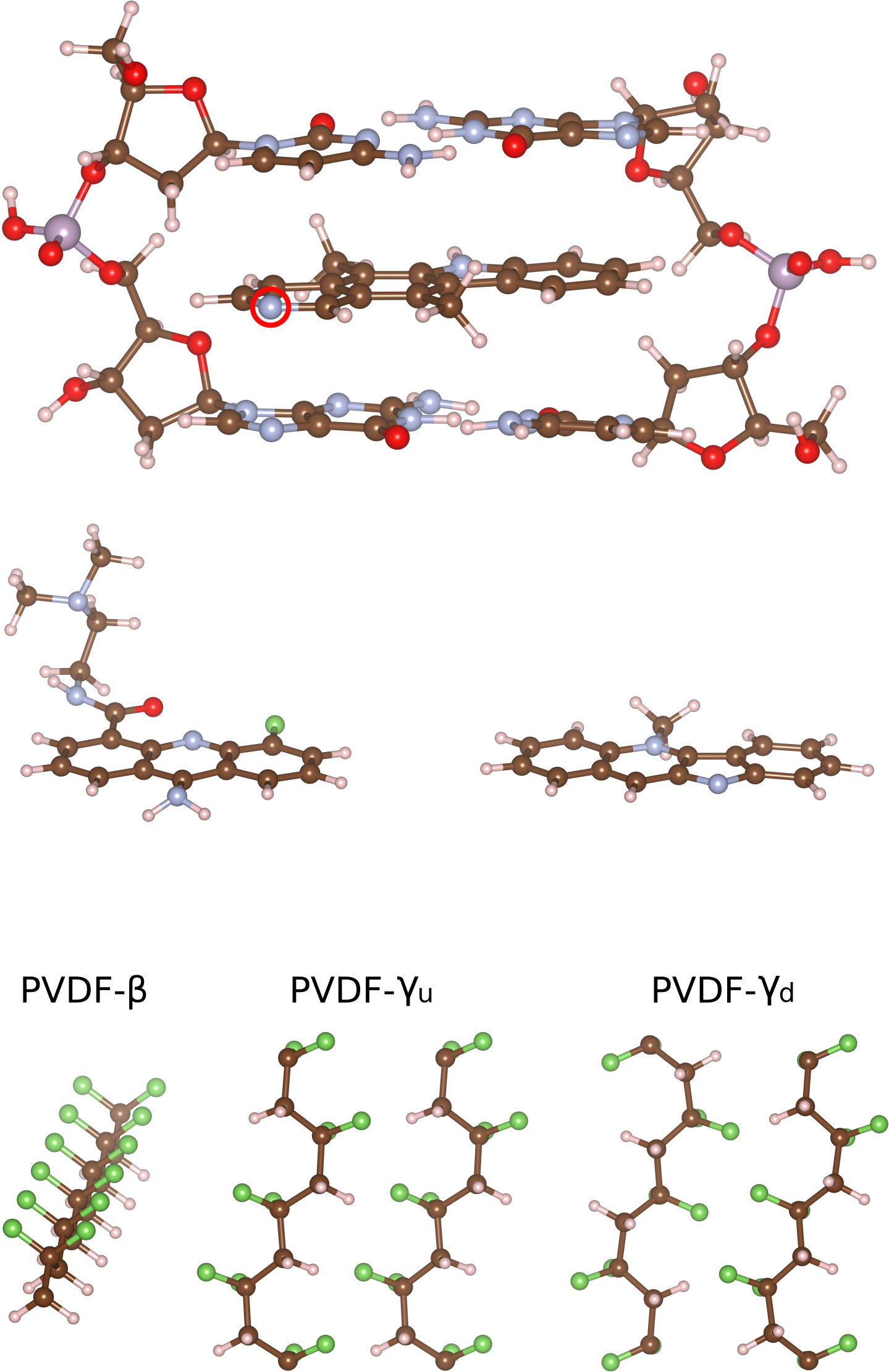}
\caption{Structure model for DNA intercalation of the molecule with PDB code 1Z3F 
(also called ellipticine, top panel), 
alternative intercalants with PDB codes 1DL8 and 1K9G (the latter also called cryptolepine, middle panels), and the ferro-electric $\beta$ as well as $\gamma_{u/d}$ forms of PVDFs (bottom row of panels). Brown, red, big/small blue, green, and white spheres identify C, O, P/N, F, and H atoms. Our CX tests are based on 
Ref.\ \onlinecite{harding2020}
reference geometries and CCSD(T) energies and we use their notation: The DNA-structure model shown in the top panel is denoted `B' and involves a protonated backbone; There are also results for a model `A' that omits the
backbone. The intercalant in the top 
panel (1Z3F, henceforth denoted `3') is also studied in a  charged form 
    `3$^+$' where a proton is added at the
    nitrogen identified by a red circle.
The alternative intercalants shown in the middle left and right panels, 1DL8 and 1K9G, are denoted `2' and `1' respectively \cite{harding2020}. Finally, the
    PVDF forms (bottom row) are studied
    in crystals forms, see Figure \ref{fig:PVDFbetaStruct} below, and
    Table \ref{tab:PVDFgamma}.}
    \label{fig:softstructures}
\end{figure}

\subsection{Biochemistry testing and organic ferroelectrics}

The study of DNA fragments and of their assembly from building blocks is a rich research field. It is a goal of 
the overall vdW-DF method and long-term research program to realize accurate computationally effective studies of structure, of defects and intercalation
\cite{harding2020}, of fluorescence-marker base substitutions \cite{SaBoLi07,WrFuDu17},
as well as of molecular dynamics to
determine the entropic effects \cite{Joost16}. All of these problems are interesting
for biochemistry in their own right. 

Meanwhile, there is potential health-technology benefit from 
realizing 
flexible materials (polymers) with a large polarization response \cite{Ruan2018}. One can, for example, envision incorporation in bandages to allow a simple electric detection of swelling 
associated with infection. Such indirect
detection could reduce the need for traditional, 
periodic visual inspections. This idea, however, hinges on 1) The  possibility of synthesizing a ferroelectric polymer with a sufficient per-monomer polarization response, and 2) 
Achieving a sufficient polymer crystal 
ordering so that the local response also serves to define a sufficiently large net electric-signal output (in connection with deformations).
Consistent vdW-DF calculations cannot directly
help synthesis, but they can predict the 
structure of perfect crystals and then 
assert if the ideal polarization response
is sufficiently large, for any given soft-ferroelectrics candidate.

We shall illustrate this for the polyvinylidene fluoride (PVDF)  system. However, since we need
to predict both structure and response, we
 first provide method validation.
Fortunately, the richness of the DNA and biochemistry field means that it offers 
many testing opportunities as we also seek a deeper account of polymer physics in technology and in 
general \cite{Lando1966,Hasegawa1972,Nakamura2001,Resta2000,kleis05p164902,kleis05p192,kleis07p100201,Ranjan2007,kleis08p205422,Ranjan2012,Itoh2014,Dong2016,Olsson17,OlHySc18,Pelizza2019}.  In all this soft-matter testing and application
work we use the ONCV-SG15 
PPs \cite{ONCV,sg15}  at 160 Ry energy cutoff
in \textsc{Quantum Espresso}. We also ensure an electrostatic mono- and dipolar
decoupling \cite{Makov} in
large cubic unit cells.

Figure \ref{fig:softstructures} schematically 
shows a class of DNA intercalation problems (top and middle rows) that we use to 
validate accuracy of the consistent-exchange CX version 
before leveraging it on the application, the ferroelectric PVDF polymer (bottom panel). 
In fact, we include two method-validation checks for soft-matter performance, namely a selection of benchmark sets from the GMTKN55 suite \cite{gmtkn55} and the DNA intercalation problems. Both focus on the CX energy description and thus supplement prior documentation of CX performance for structure and phonons in polyethylene and in oligoacene crystals \cite{RanPRB16,BrownAltvPRB16,Olsson17,OlHySc18,HyJiSh20}.

The GMTKN55 is a suite of benchmarks of broad molecular properties that also  contain a range of DNA-relevant benchmark sets. 
For a simple test of the CX, CX0P and AHCX performance on molecules
it is relevant to consider the 
S66  set within the 
GMTKN55 suite \cite{gmtkn55}. It is a set that broadly reflects noncovalent interactions.
 More biochemistry-focused checks can be extracted by also computing
mean-absolute deviations of our XC functional descriptions (relative
to coupled-cluster CCSD(T) values \cite{gmtkn55}) for the PCONF21
set of peptide conformers, the Amino20x4 set of amino-acid interaction
energies, the UPU23 set of RNA backbone conformer energies, the
SCONF set of sugar conformers, and the WATER27 set.
The testing setup is similar to what is used in Ref.\ \onlinecite{AHCX21} but here included 
also the WATER27 benchmarking set by computing 
the energy of the OH$^-$ ion in a smaller 12 {\AA} cubic cell. This allows us to circumvent
adverse convergence impact of self-interaction errors \cite{BurkeSIE,AHCX21}.

Additionally, we test by using a recent paper 
on DNA intercalation motifs \cite{harding2020}. The study identifies a set of relevant frozen coordinate geometries, Fig.\ \ref{fig:softstructures}, for which it
also provides coupled cluster CCSD(T) reference energies. We use those energies for a test of the CX performance, because the model circumvents the need for a detailed study of the effects of counter ions and water. 

The basic idea is to consider two models of a DNA base-pair segment, namely one where the backbone is protonated  (effectively placing one extra electron per phosphor group on the back bone structure), and one where the back-bone is further eliminated. There are reference energies (and structures) for  both models with 3 intercalants; In addition, there are also reference energies for a variant, where the 
intercalant is itself protonated.

For our soft-matter application study, we characterize primarily the $\beta$ crystalline form of the PVDF system, while also comparing to the
so-called $\gamma$ forms, Fig.\ \ref{fig:softstructures}. We predict the relaxed structure and ferroelectric response of perfect crystals of $\beta$- and $\gamma$-PVDF, while comparing to experiments
where possible. Both forms can be synthesized but,
to the best of our knowledge, large single-crystal
 samples, with long-range order, do not yet exist. Theoretical predictions are thus motivated and we 
here seek to provide primarily a CX characterization in a $2 \times 2 \times 8$ Monkhorst-Pack grid sampling of the Brillouin zone. 

We compute the polymer-crystal cohesive energy,
\begin{equation}
E_{\rm coh}^{\rm pol}(a,b,c) = E_{\rm crystal} -2 E_{\rm monomer}
\label{eq:cohcrys}
\end{equation} 
for a series of unit-cell lattice constants in the $\beta$-PVDF form as well as for the motifs of the $\gamma$-PVDF form.
Here again we use the stress-vdW-DF implementation in \textsc{Quantum Espresso}. We also navigated the stiffness problem that results because 
variation in the polymer-direction $c$ reflects variations in covalent bonds within a single PVDF strain, while variations in the $a$ and $b$ lattice constants reflect the noncovalent inter-polymer-strain bindings.

In practice, we proceed as follows for a characterization of
the $\beta$ phase. For a range of fixed $c$ lattice-constant choices we optimize the structure using 
forces and stresses in a \textsc{Quantum Espresso}
variable-cell calculation that 
exclusively permits $a$ and $b$ structural relaxations.
We then use a fourth-order polynomial
fit to identify the minimum $c_0$ of the 
resulting  $E_{\rm coh}^{\rm pol}(c_0,a(c_0), b(c_0))$ variation.
This identifies the optimal structure as well as the
actual crystal cohesion (in a given functional). The relaxation of PVDF in the $\gamma$-phase is carried out for the same PPs, but using a $2 \times 2 \times 8$ $k$-point grid.

Finally, an additional fourth order polynomial is used 
to define a contour plot of  the $E_{\rm coh}^{\rm pol}(c_0,a, b)$ 
variation in the soft $a$ and $b$
unit-cell directions \cite{EleniSchroder,kleis07p100201,Olsson17}. 
This is done to give a representation of the PES,
noting that the $c_0$-lattice constant
exhibits only moderate changes with the 
XC functional choice. As such, this
two-dimensional PES description
offers us a chance to identify 
overall trends of functional impact
on the PVDF characterization.

\section{Hard-matter examples}

\subsection{Cubic perovskite properties and stability}

\begin{table}[tbp]
\caption{ 
Comparison of equilibrium lattice constant $a$ (as obtained from a Birch-Murnaghan fit to DFT calculations in the BO approximation), 
frequencies of the lowest (and possibly soft) mode at the $\Gamma$- and $R$-point, 
Landau-model expansion parameters ($\bar{\kappa}, \bar{\alpha}$, see Eq. \eqref{eq:Landau-expansion}),
characteristic life-time ($\tau_R$) of the phonon trapped in the double well potential, elastic coefficients, and
the high-frequency dielectric constant $\eps_\infty$. We list results 
computed in regular and hybrid functionals having either a semilocal- (PBE and HSE) or a truly nonlocal-correlation (CX and CX0P) description; LDA results and experimental values are  included for reference. 
All results are obtained in a cubic cell at the native BO lattice constants $a$, except $\bar{\kappa}$, $\bar{\alpha}$ and $\tau_R$ which are obtained from the set of PES results, i.e., 
computed by deforming a 40-atom unit cell. 
Imaginary values in $\Gamma_{15}$ or $R_{25}$ reflect a possible or incipient  instability in the description of the cubic cell by that functional. Corresponding
identifyer `\emph{soft}' in the experimental column reflects the observation of a 
low-temperature phase transition.
}
\label{tab:main_table}
  \begin{ruledtabular}
  \begin{threeparttable}[t]
    \begin{tabular}{llrrrrrr}
\textbf{BaZrO$_3$}			&	Exper. 			&	LDA		&	PBE		&	CX		&	HSE			&	CX0P	\\ 
\hline
$a$ [Å]			            &	4.188\tnote{a}	&	4.160	&	4.237	&	4.200	&	4.200		&	4.183	\\
$\Gamma_{15}$ [meV]	        &	15.2\tnote{b}	&	13.04	&	11.97	&	13.70	&	13.47		&	14.85	\\
$R_{25}$ [meV]		        &	5.9\tnote{a}	&	$i7.17$	&	2.30	&	$i2.53$	&	6.12		&	5.49	\\
\hline
$\bar{\kappa}$ [meV/Å$^2$u] & 8.51\tnote{a}& $-$12.5	&	1.27    &	$-$1.30	&	8.54 		&	6.68	\\
$\bar{\alpha}$ [meV/Å$^4$u$^2$]	&	-	&	2.55	&	2.03	&	2.19	&	2.32		&	2.3	\\
$\tau_R$ [$10^{-12}$ s]     &        -  &   6.0     &   -       &   -    &   -           &    -\\ \hline
$C_{11}$ [GPa]		        &		282\tnote{c}/332\tnote{d}		&	348		&	290		&	324		&	298			&	340	\\
$C_{12}$ [GPa]	            &		88\tnote{c}		&	88		&	79		&	89		&	84			&	87	\\
$C_{44}$ [GPa]		        &		97\tnote{c}/97\tnote{d}		&	90		&	85		&	88		&	94			&	97	\\
$\eps_\infty$               &	    4.928\tnote{e}	&    4.92	&	4.88	&	4.88	&	4.25		&	4.32	\\
\hline 
\textbf{SrTiO$_3$}			&	Exper.			&	LDA		&	PBE		&	CX		&	HSE			&	CX0P	\\ 
\hline
$a$ [Å]			            &	3.905\tnote{f}	&	3.862	&	3.942	&	3.905	&	3.900		&	3.880	\\
  $\Gamma_{15}$ [meV]	    &	\emph{soft}		&	7.16	&	$i17.34$	&	$i6.69$	&	$i15.52$		&	$i1.38$	\\
$R_{25}$ [meV]	            &	\emph{soft}		&	$i11.15$	&	$i8.34$	&	$i8.88$	&	$i3.29$		&	$i5.56$	\\
\hline
$\bar{\kappa}$ [meV/Å$^2$u]         &	\emph{soft}		&	$-$29.2	&	$-$16.3	&	$-$18.4	&	$-$3.8		&	$-$8.9	\\
$\bar{\alpha}$ [meV/Å$^4 $u$^2$]	&	-	&	3.36	&	2.84	&	3.07	&	3.17		&	3.34	\\
$\tau_R$ [$10^{-12}$ s]             &        -  &   14400   &   19      &   36      &   0.28        &   0.58 \\
\hline
$C_{11}$ [GPa]	            &	318\tnote{g}	&	381		&	314		&	348		&	361			&	337		\\
$C_{12}$ [GPa]	            &	103\tnote{g}	&	109		&	99		&	102		&	112			&	101		\\
$C_{44}$ [GPa]	            &	124\tnote{g}	&	118		&	111		&	115		&	128			&	123		\\
$\eps_\infty$	            &				-	&	6.34	&	6.335	&	6.30	&	5.076		&	5.202	\\
\end{tabular}
     \begin{tablenotes}
		\item[a] Low temperature neutron measurements extrapolated to 0 K, Ref.~\onlinecite{PeGrRo20}
        \item[b] Ref.~\onlinecite{Helal2015}
		\item[c] Sound velocity measurements at 25\textdegree C, Ref.~\onlinecite{Gor1998}
		\item[d] Brillouin scattering at 93~K, Ref.~\onlinecite{HelMor2021}
		\item[e] Ref.~\onlinecite{AkbKorKia2005}  
		\item[f] High-resolution X-ray diffraction at room temperature, Ref.~\onlinecite{SchKwaSch2012}
		\item[g] Sound velocity measurements. Values extracted at 273 K, Ref.~\onlinecite{BelRup1963}
     \end{tablenotes}

  \end{threeparttable}
  \end{ruledtabular}
\end{table}

Table \ref{tab:main_table} reports a summary of the
BO characterizations of \bzo\ and SrTiO$_3$ as
computed using LDA, PBE, CX, HSE, and CX0P. The top half of the table supplements our previous \bzo\ characterization \cite{grwahy20} 
by here focusing on the impact that functional descriptions 
have on properties as described on the native BO
lattice constants. This \bzo\ data also provides
a reference for the SrTiO$_3$ characterizations that we include in the
bottom half of the table.

For SrTiO$_3$ we find that LDA provides
the best characterization of the high-frequency dielectric constant.
For the elastic-constant description we find that it is CX and
CX0P that delivers the most accurate description overall; 
The CX characterizations are helped by
having an essentially perfect 
line up between the optimal 
CX BO lattice
constant and experiments.

For the AFD  $R$-mode and for the $\Gamma$ modes
of \sto, we find soft modes for all functionals (except in the LDA-$\Gamma_{15}$ description) in their description of the simple-cubic phase.
Our PBE and HSE results for the soft $\Gamma_{15}$
mode are in fair agreement with the values
(PBE at $i$14 meV and HSE at $i$9 meV) reported in
Ref.\ \onlinecite{WahVogKre09}, considering
that we use a considerably larger energy cutoff.

The \sto\ results for the $R_{25}$ mode identify incipient instabilities, that is, they corresponds to possibilities
for actual structural transformations, in all XC-functional  cases. We recall that \sto\ indeed undergoes an AFD-type 
transition to a tetragonal phase at 105 K \cite{MulBer1971}, corresponding to 
compensating rotation of the
oxygen cages, Fig.\ \ref{fig:PerovStruct}.
Modeling based on a good XC functional should be consistent with this observation. Unlike the \bzo\ case, there are no low-temperature experimental data for the cubic phase but we can discuss the $T\to 0$ stability for would-be cubic \sto.

\begin{figure}[htpb]
	\includegraphics[width=0.99\columnwidth]{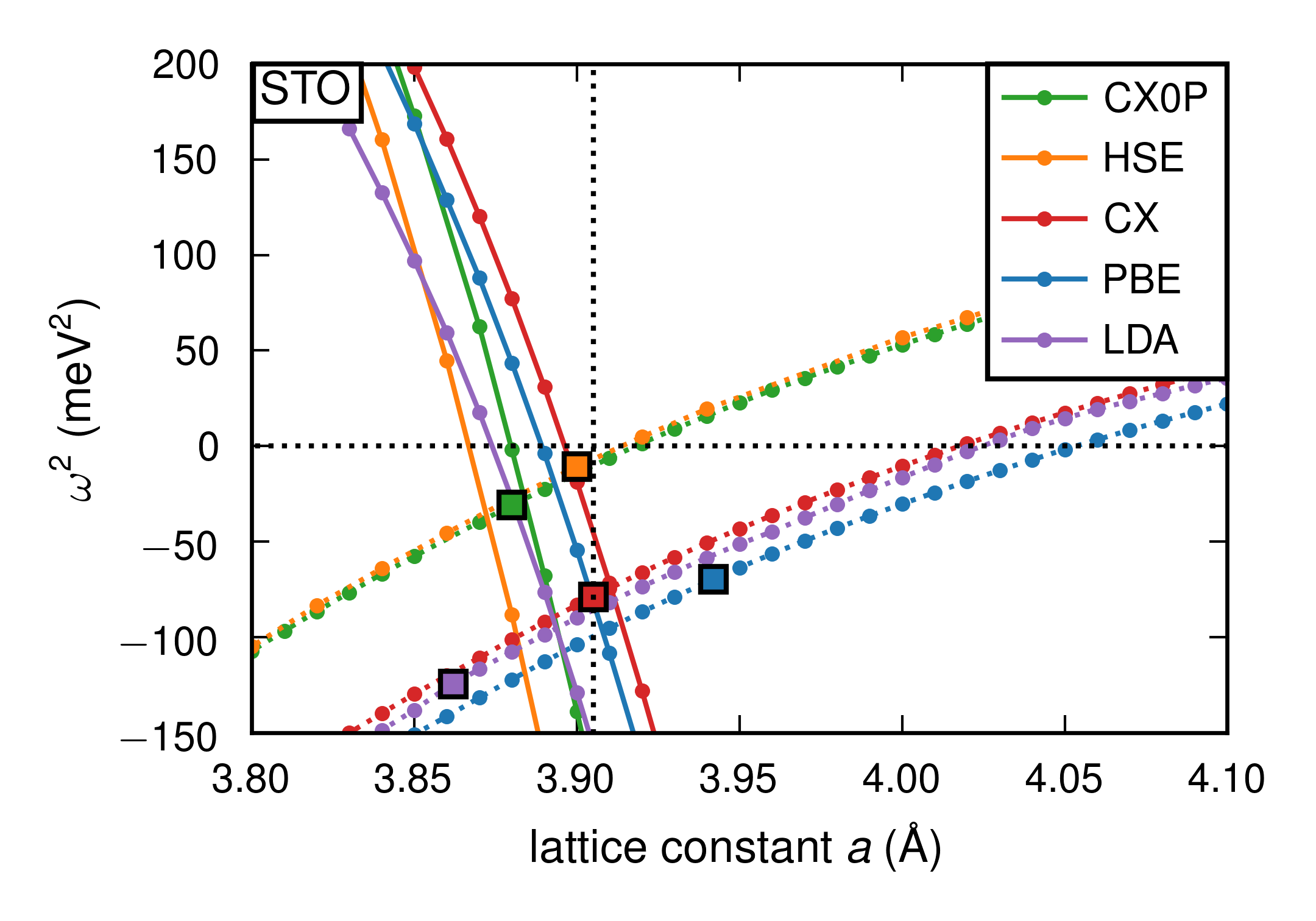}
	\caption{Evolution of the ferroelectric $\Gamma_{15}$ mode (solid) and anti-ferrodistortive $R_{25}$ mode (dashed) as function of lattice constant.
    \label{fig:frequency-stabilitySTO}
	}
\end{figure}

Fig.\  \ref{fig:frequency-stabilitySTO} shows calculations of the potentially unstable $R$ AFD and $\Gamma$ modes in cubic \sto\ as a function of assumed lattice constants for the set of XC functionals LDA, PBE, CX, HSE, and CX0P.
The figure reports frequency-square values, $\omega^2$ (as computed in \textsc{phonopy}),
reflecting a quadratic expansion in deformations in the cubic structure.
The vertical dashed line shows the experimental lattice constant at room temperature while the dashed horizontal line is set at a zero-$\omega^2$
value to delineate stability with a 
potential for instability. This is the measure
that would apply if we could ignore zero-point
energy dynamics of, for example, the AFD modes
(for which the $\omega^2$ values increase with the
lattice constant). The large squares show
the position of the optimal lattice constant
for each functional. The set of curves shows
the frequency-squared values $\omega^2$ of the
AFD $R$ and of the ferroelectric-deformation $\Gamma$ modes. These values are obtained in
finite-difference \textsc{phonopy}
calculations, for each of the functionals, as a function of the assumed cubic lattice constant.

We find that CX provides the best structural 
description in the sense that CX has both a highly 
accurate lattice constant description and that  
it's predictions for the AFD phonon behavior is consistent with the experimentally observed 
\sto\ phase transition, as argued in detail below.

On the one hand, the HSE delivers a description that is close on the lattice constant, but on the 
other, the HSE-based \textsc{phonopy} determination of the $R$-mode $\omega^2_R$ value (at the  optimal structure) is barely negative; 
Quantum fluctuations are expected to easily
compensate this small potential for an AFD-driven 
phase transition. Worse, the HSE $\Gamma$-mode $\omega_\Gamma^2$ value rapidly decreases beyond
the BO lattice constant so that
HSE instead suggests a ferroelectric \sto\ behavior, again in conflict with observations.

Use of CX0P XC input gives $\omega^2_R$ values which are more negative than those for HSE at the native CX0P lattice constant. It also yields 
a modal description with an improved resistance towards a $\Gamma$ mode (ferroelectric) instability, Fig.\ \ref{fig:frequency-stabilitySTO}. However, CX0P is still in conflict with experimental observations: When the CX0P input (for $\omega_R^2$) is adjusted to the experimental lattice constant (vertical dashed line), it gives again only a very weak AFD-mode instability. Use of CX0P does not lead to the prediction of an actual low-temperature transformation in \sto. Still, it is interesting to contrast the CX0P stability analysis with that
of CX.

Table \ref{tab:main_table} lists 
the parameters $\bar{\kappa}$ and $\bar{\alpha}$ that characterize the fourth-order fits
to DFT results in the 5 functionals we use to
model \sto. We note that large negative $\bar{\kappa}$ values correspond to deep double-well potentials 
which occur for LDA, CX (and to a similar extent for PBE). However, the total-energy variation (and hence the potential for the effective AFD-model Hamiltonian, Section II.D) is shallow 
for CX0P. We also note that the quality and the consistency of the Landau modeling (given by $\bar{\kappa}$ and $\bar{\alpha}$) is confirmed by checking the \textsc{phonopy} results,
Table \ref{tab:main_table}, using the 
approximation $\omega_R^2\approx \bar{\kappa}$.

In Table \ref{tab:main_table} we also list the
tunneling times $\tau_R$ as extracted for the 
set of functionals, using Eq.\ (\ref{eq:modaltimeDef}).
We find  that LDA is characterized by very long (ns) dwell or tunneling times, while 
CX and PBE give moderately long times (36 and 19 ps). Finally, the table reports that CX0P and HSE are characterized by short dwell times (less than one ps). With an assumed dephasing time on the order of 1 ps, our simple stability
model suggests that a CX (CX0P) description 
predicts (does not predict) an actual distortion at $T\to 0$.

In summary, we find that CX yields a better \sto\ characterization than CX0P, when also considering the stability question. This is in contrast to the
case of \bzo, where CX0P performs better than CX for a broad study of properties. 
The perovskites is an example where more
studies are needed to assert when we can 
systematically leverage hybrid advantages 
in combination with truly nonlocal correlations
(as in CX0P and AHCX).

\subsection{Magnetic elements and BiMnO$_3$}

\begin{figure}
\includegraphics[width=1\linewidth]{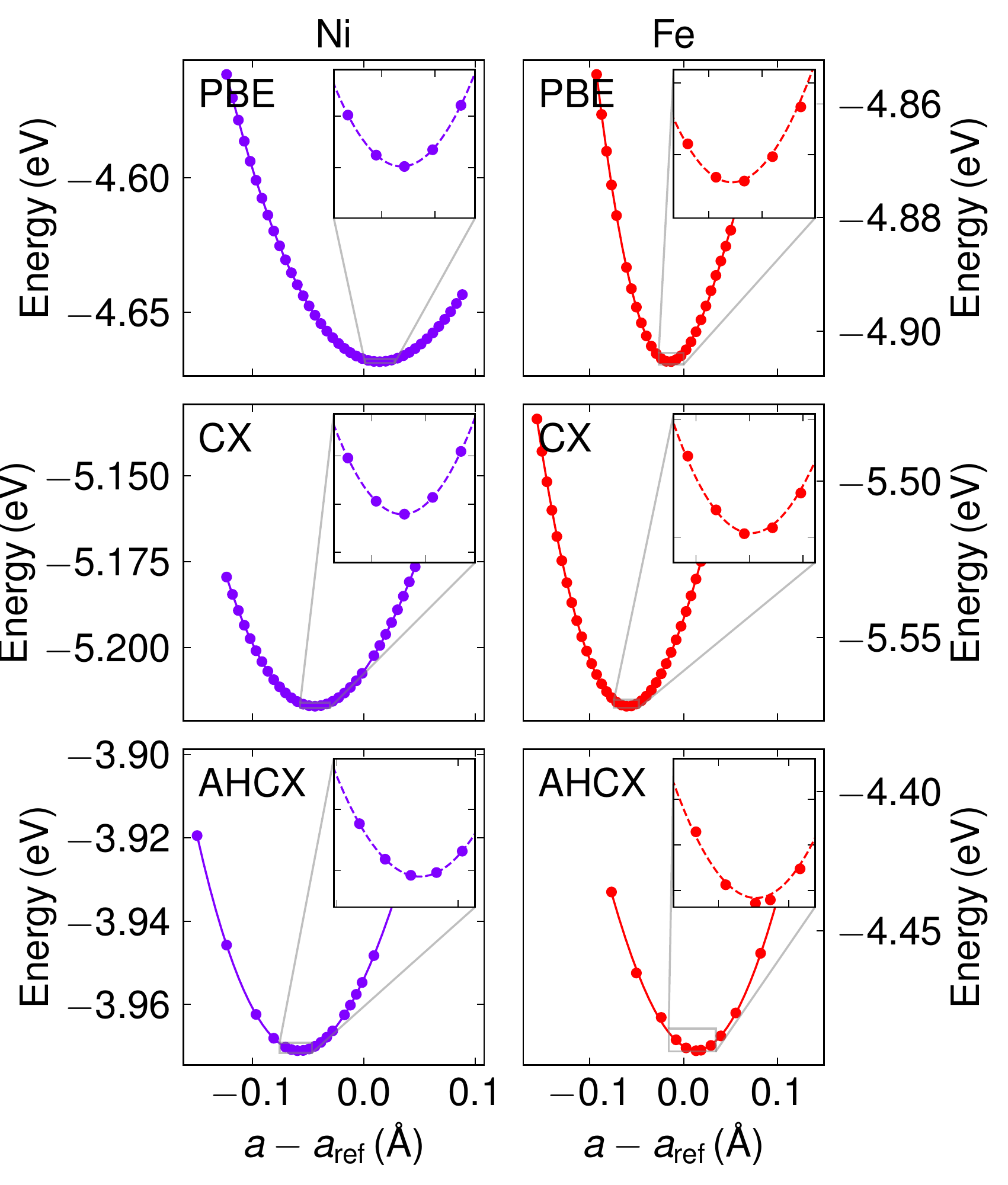}
\caption{Cohesive-energy variation with lattice constant
for Ni and Fe;  See Table \ref{tab:BMvsPoly} for a summary of
bulk structure characterization. The dots are the calculated values, the line is a fitting to a fourth order polynomial, 
Ref.~\onlinecite{EleniSchroder}. The curves are plotted relative
to the experimental lattice constant $a_{\rm ref}$, 
back-corrected for zero-point and thermal vibrational 
energy effects.}
   \label{fig:bar_partial}
\end{figure}

\begin{table}
\caption{Structure properties of magnetic elements Ni and Fe obtained in spin-vdW-DF stress calculations, where practical, and checked by the polynomial fitting. The table compares for PBE, vdW-DF-cx and
the associated range-separated vdW-DF-ahcx
hybrid, the lattice constant $a_0$, cohesive energy $E_{\rm coh}$ and bulk modulus $B_0$ against back-corrected experimental values, i.e.,
adjusted for zero-point and thermal vibrational lattice effects.}
\begin{ruledtabular}
\begin{threeparttable}
\begin{tabular}{l|ccc|l}
     & PBE         & CX                   & AHCX                    & {Exper.*}$^a$        \\
\hline           
\textbf{Ni}                  &             &             &          &                       \\
$a_{0, \rm{fit}} $ [Å]     & $3.524$     & $3.466$     & $3.452$  & ${3.510}$      \\
$a_{0, \rm{stress}} $ [Å]  & $3.524$     & $3.466$     & $-$      & ${}$           \\
$E_{\rm{coh}}$ [eV]          & $4.668$     & $5.217$     & $3.971$  & ${4.477}$      \\
$B_0$ [GPa]                  & $197.0$     & $226.3$     & $227.1$  & ${192.5}$      \\
\hline
\textbf{Fe}                  &              &            &          &                \\
$a_{0, \rm{fit}} $  [Å]    & $2.839$      & $2.795$    & $2.868$  & ${2.855}$      \\
$a_{0, \rm{stress}} $ [Å]  & $2.840$      & $2.796$    & $-$      & ${}$           \\
$E_{\rm{coh}}$ [eV]          & $4.905$      & $5.572$    & $4.498$  & ${4.322}$      \\
$B_0$  [GPa]                 & $158.1$      & $216.1$    & $184.9$  & ${168.3}$      \\
\end{tabular}
\label{tab:BMvsPoly}
\begin{tablenotes}
		\item[$a$] Ref.~\onlinecite{Lejaeghere14} 
\end{tablenotes}
\end{threeparttable}
\end{ruledtabular}
\end{table}

Figure \ref{fig:bar_partial} compares results in PBE, CX and AHCX for
the cohesive energy for Ni (left column) and Fe (right column). The panels track the overall cohesive energy variation with the
lattice constants $a$ to identify the optimal native 
BO structure and associated cohesive energy 
for these magnetic elements. 
We compute sets of (PBE as well as) CX and AHCX cohesive energy values at a set of frozen lattice constants $a$ (dots) and we fit 
those energy variations to fourth-order polynomials (solid curves); 
The inserts validate the consistency of these fits (showing that they go through the computed minima). 
We compare our results for Ni and Fe to experimental values that have been back-corrected to account for zero-point energy and temperature vibrational effects.
Our results and our approach yield both a comparison 
 of CX and AHCX performance
and an opportunity for a simple test of our new spin-vdW-DF stress description in the CX case.

In Table \ref{tab:BMvsPoly} we summarize the results of our magnetic-element structure characterizations.  We see that while CX performs very well on
structure and cohesion, across the set of nonmagnetic transition metals
\cite{Gharaee2017}, it leads to a slight underestimation of the Ni lattice constants and a significant underestimation for Fe. However, we also 
find that the new AHCX partially corrects the Fe description.

Of direct interest for the here-presented theory and code work, Table \ref{tab:BMvsPoly} also shows that for CX (where we can use the spin-vdW-DF stress description in our updated \textsc{Quantum Espresso} code),
there is a near-perfect alignment of the variable-cell Fe and Ni 
descriptions and the results obtained through polynomial
fitting for the minima, Fig.\ \ref{fig:bar_partial}. We find our
new spin-vdW-DF stress description validated.

\begin{table}[h]
\caption{\label{tab:BMO} Ground-state structure of the ferromagnetic and ferroelectric BiMnO$_3$. The volume $\Omega$ is reported per formula unit. For structure we compare to experiments at 20 K, Ref.~\onlinecite{Attfield-BMO}. For the length of the polarization vector $|\bm{P}|$, we compare to experiments obtained at 5 K, Ref~\onlinecite{Hyoungjeen2011}. }
\begin{ruledtabular}
\begin{threeparttable}
\begin{tabular}{lcc}
                             &    CX      
                             & Exper.      \\
\hline                                                                           
$a$ [Å]                      &   $9.49$   
                             &  $9.52$  \\
$b$ [Å]                      &   $5.57$   
                             &  $5.59$  \\
$c$ [Å]                      &   $9.62$   
                             &  $9.84$  \\
$\alpha$ [$^\circ$]          &   $90.1$   
                             &  $90.0$    \\
$\beta$ [$^\circ$]           &   $109.3$  
                             &  $110.6$   \\
$\gamma$ [$^\circ$]          &  $91.62$   
                             &  $90.0$   \\
$\Omega$ [$\rm{\AA}^3$]      &  $59.93$   &  $61.40$   \\

$|\bm{P}|$ [ $\mu {\rm C} / {\rm cm}^2$ ]     &   $36$  & $23$ \\
\end{tabular}
\label{tab:BiMnO}
\end{threeparttable}
\end{ruledtabular}
\end{table}

Table \ref{tab:BMO} lists experimental observations on BiMnO$_3$, obtained using both X-ray diffraction and neutron diffraction at 5-300 K, \cite{Hyoungjeen2011}. The system has a ferromagnetic ordering. Accordingly, the data represents an additional  route for us to illustrate use of the new spin-vdW-DF stress formulation, while also permitting a test of the accuracy of spin CX 
calculations.

We find that CX 
is accurate on the
angles and side lengths, especially for the  $a$ and $b$ lattice constants that define the basis plane in the schematics, Fig.\ \ref{fig:structs}. Accuracy in the description of this in-plane structure is important since this plane contains the main BiMnO$_3$ ferroelectric response. Overall, we find that the spin-vdW-DF stress description is both numerically stable and works in the sense that it reliably predicts an optimal structure even in more complex magnetic perovskite structure cases.

\begin{figure}
\includegraphics[width=0.9\columnwidth]{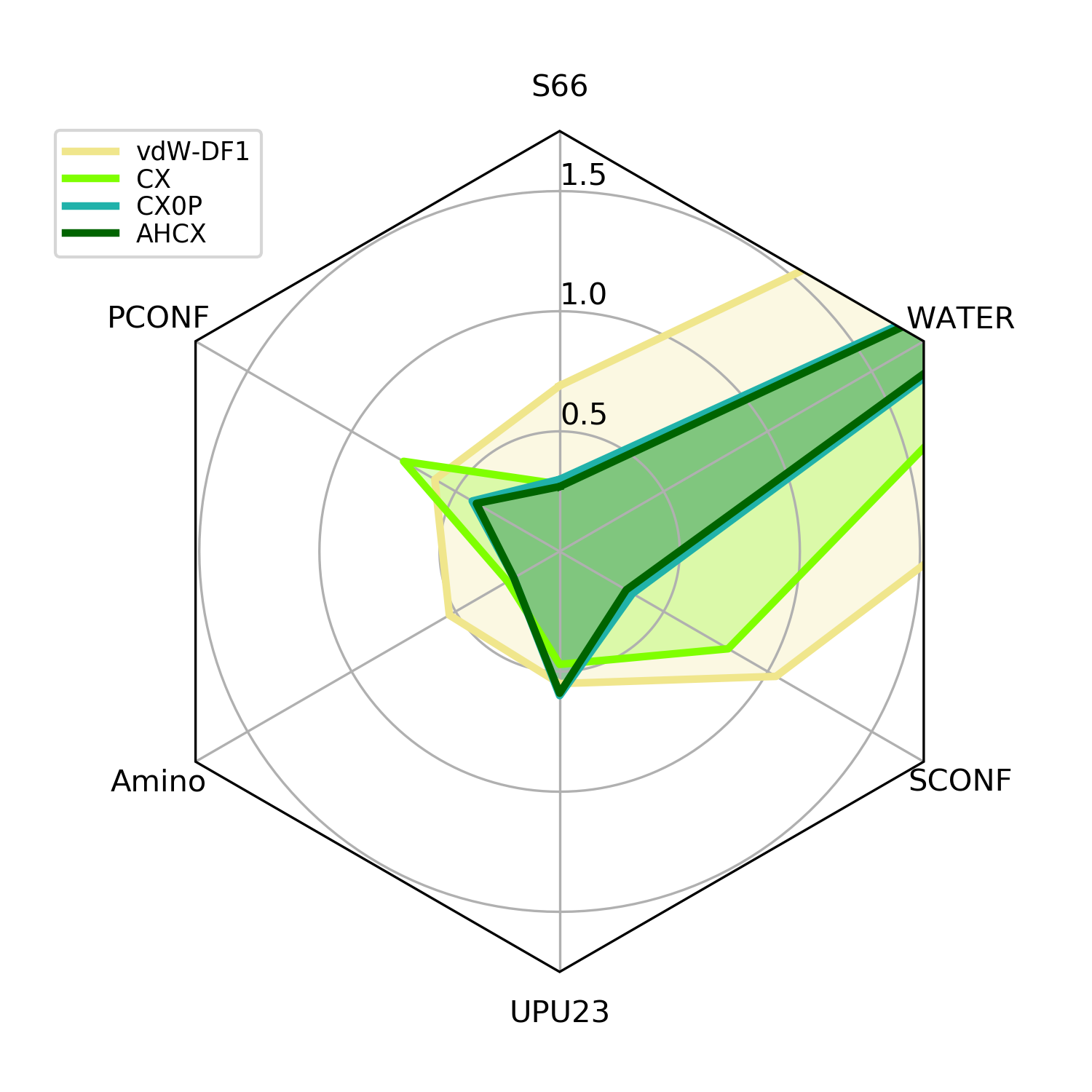}
\caption{Performance assessment of the consistent-vdW-DF tool chain, CX-CX0P-AHCX, on 6 benchmark sets \cite{gmtkn55} of key bio-chemistry relevance, comparing relative molecule energies in: `S66' (a balanced set), PCONF (tri- and tetra-peptide conformers), 
Amino20x4 (amino-acid conformers), 
UPU23 (RNA backbone conformers), 
SCONF (sugar conformers), and 
WATER27 (binding energies of water complexes, including some cases with 
proton transfers). We show
mean-absolute deviation (MAD) values (expressed in kcal/mol)
from quantum-chemistry CCSD(T) reference calculations \cite{gmtkn55}, noting that performance of the unscreened
hybrid CX0P (turquoise curve) is almost on par with that of the range-separated hybrid AHCX (dark green curve). We also include a 
performance overview of the original vdW-DF1 version \cite{Dion04,Dion05} that shares the nonlocal-correlation energy formulation.
The WATER27 benchmarking on the consistent-vdW-DF tool chain yields slightly larger MAD values:  2.89, 2.84, and 2.80 kcal/mol for CX, CX0P, and AHCX).
}
\label{fig:BioRadar}
\end{figure}

\section{Soft-matter examples}

\begin{table}
\caption{Comparison of CX intercalation energies, in kcal/mol, with 
CCSD(T), B3LYP-D3, and HF-3c results from Ref.~\protect\cite{harding2020}. All structures are computed at the
experimentally motivated reference geometries, considering a  double-basis
set DNA-model without or with a (protonated) backbone, labeled `A' and `B', respectively \cite{harding2020}. The CCSD(T) reference energies exist
for three intercalating molecules, here labeled `1', `2', and `3', allowing also for a protonated molecule variant, denoted `3$^+$'.
The comparison with dispersion-corrected 
B3LYP results are for results obtained
with the def2-TZVP basis set, using the
Becke-Johnson damping function \cite{becke07p154108} on the semi-empirical Grimme-D3 correction term
\cite{grimme3,gmtkn55}. The mean absolute deviations (MAD) from the CCSD(T)-calculations
are given for CX, B3LYP-D3, and HF-3c.}
\label{tab:DNA-intercalants-CCSD}
  \begin{ruledtabular}
  \begin{threeparttable}[t]
\begin{tabular}{lcccc}
System & CX & CCSD(T) & 
B3LYP-D3 & HF-3c\\
\hline
1 A   & -39.65  & -41.99  & -41.1& -40.3\\
2 A   & -39.51  & -39.52  & -39.9& -35.7 \\
3 A   & -34.09  & -34.57  & -34.4& -32.4\\
3$^+$A&-47.04  & -47.74  & -47.9& -44.5\\
\hline
1 B   & -43.60  & -45.44  & -45.2& -44.3\\
2 B   & -45.25  & -45.25  & -45.7& -42.5\\
3 B   & -39.86  & -39.39  & -40.2& -39.1\\
3$^+$B&-61.86  & -62.55  & -63.8& -61.6\\
\hline 
MAD  & 0.82 & - & 0.54 & 2.00 \\
\end{tabular}
  \end{threeparttable}
  \end{ruledtabular}

\end{table}

\subsection{DNA-related tests of vdW-DF-cx accuracy}

Figure \ref{fig:BioRadar} shows a performance comparison 
for the CX-based consistent-vdW-DF tool chain, on a subset of benchmarks of the GMTKN55 suite that probes broad molecular
properties. The benchmark sets were identified in Section III
above and focus on bio-relevant problems, for example,
peptides, RNA back bone, and sugar conformers (noting that
sugar also defines the DNA back bone). The performance of the
consistent-vdW-DFs (CX/CX0P/AHCX) is 
strong overall on molecules \cite{AHCX21}
and very strong for most of these bio-relevant 
problems. The performance is clearly better than, for example,
that of the original vdW-DF1 version \cite{Dion04,Dion05}.

The WATER27 benchmark set constitutes a challenge for most
XC functionals \cite{gmtkn55,HyJiSh20}. Even here we
find that the consistent-vdW-DF tool chain delivers 
a robust description, with a mean-absolute deviation 
(MAD) value of 2.8 kcal/mol for AHCX and almost as good 
for the nonhybrid CX form. This suggests that CX and the
tool chain is useful for determining interaction energies
in biochemistry and, we expect, in both bio- and synthetic 
polymers. This positive test outcome 
substantiates previous results, finding a
high CX accuracy for predicting the crystal 
structure of both oligoacene and
polyethylene \cite{RanPRB16,Olsson17,OlHySc18}.

Table \ref{tab:DNA-intercalants-CCSD}  summarizes 
the additional CX benchmarking that we provide by studying intercalation energies for three nearly-planar molecules
inserted into sections of DNA, as illustrated in the top and 
middle panels of Fig.\ \ref{fig:softstructures}. 
In Ref.\ \onlinecite{harding2020} the atomic positions of 
the three intercalants, along with those of the immediate 
surrounding DNA structure, were
extracted from the Protein Database (PDB) \cite{pdb}, and the 
interaction energies were calculated using the focal point approach \cite{Csaszar1998}
for extrapolated CCSD(T) results. We here use these CCSD(T)-extrapolated results
as a benchmark. The structures used for the CCSD(T) results 
in Ref.\ \cite{harding2020} were truncated to the base pairs above and below the
intercalant, plus the part of the sugar-phosphate backbone that connects them 
(model `B', top panel of Fig.\ \ref{fig:softstructures}), or without the backbone 
(model `A').
Their models do not include solvent molecules or counter ions, and the backbones 
(model `B') or the base pairs (model `A') are passivated by protonation. 
The three intercalants have PDB codes 1K9G, 1DL8 and 1Z3F and are here (and in 
Ref.\ \cite{harding2020}) denoted `1', `2', and `3' (see top and middle panels of 
Fig.\ \ref{fig:softstructures}). 
The latter molecule includes a nitrogen atom that can be protonated and this is also 
included in the set of calculations, the protonated molecule is denoted `3$^+$'.

Table \ref{tab:DNA-intercalants-CCSD} lists the interaction energies of 
molecules `1', `2', `3', and `3$^+$' in DNA models `A' and `B', as calculated 
in Ref.\ \cite{harding2020} for CCSD(T), B3LYP-D3, and HF-3c, and calculated
here using CX. All calculations are with atomic positions fixed to those extracted 
from PDB. Inspecting the numbers in Table \ref{tab:DNA-intercalants-CCSD}, we see 
that CX yields results that are close
to those from the CCSD(T) calculations, also for the protonated molecule (`3$^+$'). This holds
whether the DNA backbone is included (model `B') or not (model `A'). 
The largest deviation for CX is seen for molecule `1' (in both DNA models),
with a 1.8-2.4 kcal/mol difference from the CCSD(T)-energies. All other deviations 
are less 
than 0.7 kcal/mol, and MAD is 0.82 kcal/mol for CX with respect to CCSD(T). This can be compared to the B3LYP-D3 deviations from CCSD(T), 
that show a slightly smaller MAD (0.54 kcal/mol) on the set of intercalate structures.
However, B3LYP-D3 (and HF-3c) are  hybrid calculations, and in plane-wave codes 
molecular hybrid calculations are found to be up to 30 times slower compared to 
regular functionals, like CX, for similar system sizes \cite{AHCX21}. 

Turning to the results for the minimal-basis method HF-3c \cite{harding2020} we see 
that the MAD value is more than double that of the CX results, at 2.00 kcal/mol. 
In other words, CX competes well with the best hybrid results of 
Ref.\ \cite{harding2020}, and offers a path to acceleration that gives improved predictability 
compared to the HF-3c minimal-basis-set approach. 

\subsection{PVDF: a ferro-electric polymer crystal}      
 
\begin{figure*}[htpb]
    \includegraphics[width=0.88\textwidth]{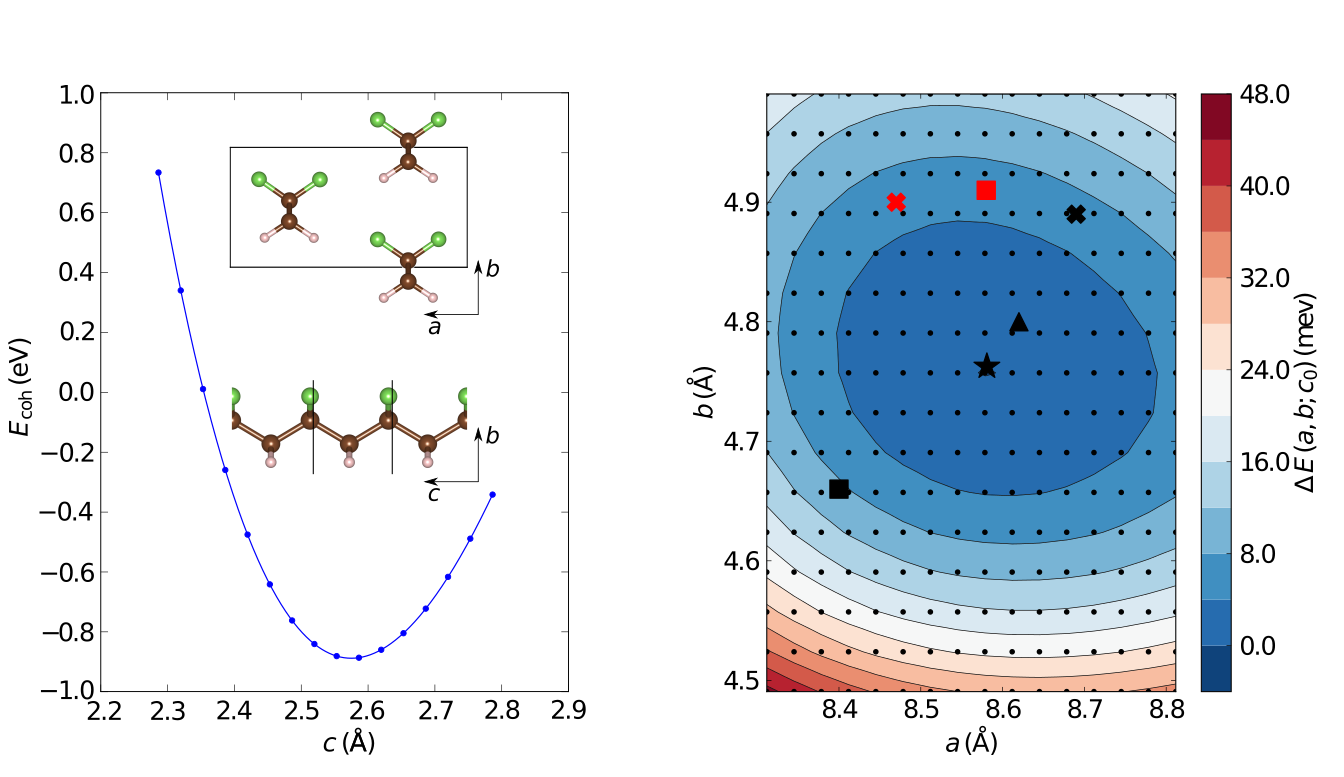}
    \caption{PVDF energy dependence on lattice parameters generated by a set of vdW-DF-cx calculations (dots). 
    The left panel shows the cohesive energy for a constrained optimization, the right panel shows contours along $a$ and $b$ generated by a triangular mesh based on the interpolated relax calculations.
    The energy landscapes are given relative to the energy of the relaxed structure (marked by the star in right panel).
    The black triangle and square are the relaxed results using vdW-DF and vdW-DF2, respectively \cite{Pelizza2019}, while the black cross shows relaxed result for PBE0 \cite{Itoh2014}.
    The red cross shows the results from X-ray diffraction at $323$ K \cite{Lando1966} while the red square is X-ray diffraction from Ref.\ \onlinecite{Hasegawa1972}.
    \label{fig:PVDFbetaStruct}}
\end{figure*}

\begin{table}[htbp]
    \caption{Calculated lattice parameters for the $\beta$-PVDF crystal. We 
    compare with literature calculational results and room-temperature experiment data.}
\begin{ruledtabular}
\begin{threeparttable}
\begin{tabular}{l|cccc}
                                & $a_0 $ [Å] & $b_0 $ [Å] & $c_0 $ [Å] & $\Omega$ [Å$^3$]   \\
\hline
PBE0 \tnote{a}                  & $8.64$     & $4.82$     & $2.64$  &  $109.2$     \\
PBE  \tnote{b}                  & $8.95$     & $5.00$     & $2.59$  &  $115.7$     \\
LDA  \tnote{b}                  & $7.97$     & $4.46$     & $2.56$  &  $ 90.2$     \\
vdW-DF1  \tnote{b}              & $8.62$     & $4.80$     & $2.60$  &  $107.5$      \\
vdW-DF2 \tnote{b}               & $8.40$     & $4.66$     & $2.60$  &  $101.8$     \\
\hline
vdW-DF1                         & $8.61$     & $4.79$     & $2.60$  &  $107.1$      \\
vdW-DF2                         & $8.38$     & $4.67$     & $2.59$  &  $101.5$     \\
CX, constrained fit             & $8.581$    & $4.763$    & $2.575$ &  $105.3$      \\
CX, vc-relax                    & $8.585$    & $4.745$    & $2.575$ &  $104.9$      \\
\hline
X-ray diffraction \tnote{c}     & $8.47$     & $4.90$     & $2.56$  &  $106.2$     \\
X-ray diffraction \tnote{d}     & $8.58$     & $4.91$     & $2.56$  &  $107.8$     \\
\end{tabular}
\label{tab:PVDF_structure}
     \begin{tablenotes}
		\item[$a$] Ref.~\onlinecite{Itoh2014} 
		\item[$b$] Ref.~\onlinecite{Pelizza2019}
		\item[$c$] Ref.~\onlinecite{Lando1966}
		\item[$d$] Ref.~\onlinecite{Hasegawa1972}
     \end{tablenotes}
\end{threeparttable}
\end{ruledtabular}
\end{table}

Figure \ref{fig:PVDFbetaStruct} shows an overview of the
structure search and potential-energy landscape for deformations of $\beta$-PVDF that we have computed using CX (as well as in vdW-DF1 and vdW-DF2). The structure is obtained using a two-stage unit-cell relaxation scheme that handles the significant stiffness-anisotropy problem in PVDF and polymer crystals in general. 

The figure also summarizes the results of 
this optimal-structure determination for $\beta$-PVDF.
The intra-polymer bonds sets the
optimal structure in $c$ direction. These are of 
covalent nature and much stronger (with 
higher strain-energy) than 
the noncovalent inter-polymer bindings that govern
the $a$ and $b$ lattice constants in the crystal. 
Accordingly we first
pick a range of possible $c$ lattice constants 
and proceed with variable-cell calculations
that exclusive relaxes the $a$ and $b$ constants 
(as well as the atomic coordinates). This gives us the $E(c)$
shown in the left panel, from which we extract the
optimal $c_0$ lattice constant as well as the rest of the
lattice constants and crystal cohesive energy.

The right panel of Fig.\ \ref{fig:PVDFbetaStruct}
shows an additional survey of the potential-energy
variation that characterizes the crystal at the optimal $c_0$
value. This mapping is obtained for CX  by varying the
$a$ and $b$ values and allowing for atomic relaxations. It is a fixed-$c_0$ projection of the ground-state energy variation. Furthermore, because the stiffness in the $c$ direction is by far the highest in $\beta$-PVDF, this projection or PES gives an impression of the energy gradients that determine the lowest-lying vibrational excitations in the $\beta$-PVDF crystal. 

Table \ref{tab:PVDF_structure} summarizes the CX structure characterization along with those obtained using vdW-DF1 and vdW-DF2 and  those found
in literature. The right panel of Fig.\ \ref{fig:PVDFbetaStruct}
provides an overview, using color symbols. We note that this is possible (even if the panel shows 
a potential-energy variation at the fixed optimal CX-$c_0$ value), because
the XC functionals are in fair agreement on the $c_0$ value.

We find that CX results obtained with the above-described
constrained-fit procedure align with those  
obtained in a constraint-free variable-cell optimization. 
However, that is true only when we start
the system description of the polymer crystal
close to the actual ground-state structure.
In Table \ref{tab:PVDF_structure} we give
an extra decimal in the reporting to facilitate 
this comparison.  

We also find that CX provides a highly accurate prediction of the optimal 
$c$ lattice constant for the $\beta$-form relative to experiments; That is, the description of 
unit-cell extension along the polymer chains is almost spot-on the
experimental observation. 
In the other directions, we find an overestimation of 
the $a$ lattice constant  but an underestimation
of the $b$ lattice constant. The PBE0 and vdW-DF1 results 
are in closer agreement with the $b$ lattice constant but there
the $c$ lattice constant is significantly 
overestimated.

\begin{table}[htbp]
    \caption{ Unit-cell lattice parameters for $\gamma$-phases of PVDF, obtained using full unit cell relaxations. The experimental values rely on a sample that contains a mixture of the up and down configurations, characterized at $T = 300$ K. The unit cell is nearly orthorhombic, with a small tilt of the along-strain axis $c$ and the $a-b$ basis plane.}
\begin{ruledtabular}
\begin{threeparttable}
\begin{tabular}{lccccc}
Functional                  &  $a [{\rm \AA}]$  &  $b [{\rm \AA}]$  &  $c [{\rm \AA}]$  & $\Omega [{\rm \AA}^3]$    & $\angle ac [^\circ]$   \\
\hline            
\multicolumn{4}{l}{$\bm{\gamma_d}$ Phase} \\
vdW-DF1                     &  $9.41$           &  $5.08$           &  $9.50$           & $454.97$                & $90.0$ \\
vdW-DF2                     &  $9.17$           &  $4.99$           &  $9.43$           & $431.90$                & $90.2$ \\
CX                          &  $9.31$           &  $5.02$           &  $9.60$           & $449.28$                & $90.0$ \\
\hline
\multicolumn{4}{l}{$\bm{\gamma_u}$ Phase} \\                            
vdW-DF1                     &  $9.60$           &  $4.98$           &  $9.31$           & $444.32$                & $96.5$ \\
vdW-DF2                     &  $9.34$           &  $4.86$           &  $9.31$           & $421.42$                & $96.0$ \\
CX                          &  $9.36$           &  $4.84$           &  $9.32$           & $421.81$                & $94.6$ \\
\hline                          
Exper.$^a$                  &  $9.67$           &  $4.96$           &  $9.20$           & $440.65$                  & $93$ \\
\end{tabular}
\label{tab:PVDFgamma}
     \begin{tablenotes}
		\item[$a$] Ref.~\onlinecite{Lovinger1981}
     \end{tablenotes}
\end{threeparttable}
\end{ruledtabular}
\end{table}

The study of the $\gamma$-PVDF crystal offers additional opportunities for a theory-experiment comparison. These crystal forms
are only nearly orthorhombic, 
characterized by a tilt angle
between the $a-b$ basis plane
and the $c$ axis that is (as for
$\beta$-PVDF) aligned with the
polymer strains. The comparison is complicated by the fact that there are two conformers, denoted $\gamma_{u}$ and $\gamma_d$, see 
Fig.\ \ref{fig:softstructures}.

\begin{table}[htbp]
    \caption{Predictions of spontaneous polarisation of the three PVDF phases, studied using different vdW-DF releases.
    All the values are presented in the unit of $\mu \textrm{C cm}^{-2}$.
    It should be noted that experimental determination of the spontaneous polarization is hard as the samples are often mixtures of both various polymorphs and noncrystalline regions. Accordingly, the same experimental range is reported for both $\gamma$ phases.}
\begin{ruledtabular}
\begin{threeparttable}
\begin{tabular}{lcccc}
Phase                       &  vdW-DF1       &  vdW-DF2          &  CX      &   Exper.           \\
\hline            
$\beta$                     &  $52.0$   &   $51.2$      &  $53.2$  &   $10^a$       \\
$\gamma_{\textrm u}$        &  $9.5$    &   $7.2$       &  $7.7$   &   $0.2-0.3^b$        \\
$\gamma_{\textrm d}$        &  $7.6$   &   $5.3$       &  $8.2$  &   $0.2-0.3^b$        \\
\end{tabular}
\label{tab:PVDF_berryphase}
     \begin{tablenotes}
		\item[$a$] Ref.~\onlinecite{Nakamura2001}
		\item[$b$] Ref.~\onlinecite{Zhao2016}
     \end{tablenotes}
\end{threeparttable}
\end{ruledtabular}
\end{table}
 
Table \ref{tab:PVDFgamma} shows the results of a vdW-DF1, vdW-DF2, and CX structure characterization, along with experimental observations for $\gamma$-PVDF \cite{Lovinger1981}. These reference 
values are obtained at room temperature in systems that are known to contain a mixture of the two motifs, i.e., both $\gamma_{u}$ and $\gamma_d$. We note that one would expect the volume $\Omega$, the tilt-angle,
and the lengths of the set of unit-cell vectors to be a linear combination of $\gamma_{u}$ and $\gamma_d$ values. Moreover, polymers are known to exhibits a significant temperature expansion \cite{RanPRB16}; A good description should be smaller than the volume measured at 300 K. 

We find that the computed structures for the $\gamma_d$ motif differ from the room-temperature experimental characterization, for all vdW-DF functionals. The tilt angle differs
and so does the predictions for the 
along-strain extension (length of $c$). The CX characterization of the $\gamma_d$ motif is, overall, not 
as close to the measurement as vdW-DF2. We note that CX is often found accurate on covalent/metallic bonding and structure, Refs.\ \cite{bearcoleluscthhy14,RanPRB16,BrownAltvPRB16,Gharaee2017,HyJiSh20} 
and that holds also for 
$\beta$-PVDF, Table \ref{tab:PVDF_structure}. 

Assuming instead that the structure
of the experimental sample is dominated by the $\gamma_u$ motif, the set of vdW-DF predictions are
closer to the data. Here, CX describes a unit-cell tilt angle that is in good agreement with the meaurement. Also, both vdW-DF2 and CX are now found to give a unit-cell description that 
is 5\% smaller than the
room-temperature measurements and the CX lattice constant 
is now in good (1\%) agreement with the measured $c$-axis extension. 

Overall, the CX is found accurate
on structure for $\beta$-PVDF and 
consistent with the mixed-$\gamma$-motif measurements.
This, in turn, makes it relevant to use CX also for predicting and comparing the polymer ferroelectric response. 

Table \ref{tab:PVDF_berryphase}
summarises this response survey. The polarization is computed using the Berry-phase approach \cite{VanKin1993,Resta1993,ResVan2007}. This is done under the assumption of having a perfect crystal at the optimal structure (as computed for each of the
vdW-DFs). We repeat that fully crystalline samples do not
yet exists and that polarization
measurements will be affected
by compensating responses arising
in different sample regions.

Hence, we provide these theoretical characterizations not as a
performance assessment but 
as an application example.
Polymer scientists cannot 
easily assert an upper limit 
on the polymer response from present PVDF measurements, but modern DFT can provide and compare, for example, among PVDF forms. Our CX calculations confirm that the $\beta$-PVDF form has the highest limit on the polarization response and may therefore eventually serve as a good organic ferroelectrics.

\section{Discussion and summary}

An overall goal of this theory, code, validation and application paper was to illustrate potential materials-theory advantages of having a tool chain of related consistent-vdW-DF XC functionals, namely
CX, CX0P, and the new AHCX. Our range-separated AHCX hybrid is very new but we have been able to include a few examples that nevertheless identify application strengths beyond those discussed
in the AHCX launching work \cite{AHCX21}. By providing a CX-based set of related XC tools, we have the option of including both truly nonlocal correlation and truly nonlocal exchange (to an increasing extent) all within the electron-gas tradition. As such it provides the same advantages for vdW-dominated problems (and hard and soft matter in general) that the PBE-PBE0-HSE chain provides in the framework of semilocal-correlation descriptions. It provides a platform for developing a systematic analysis, as the consistent-vdW-DF application range grow.

More specific goals were to upgrade the proper spin vdW-DF formulation with a stress description and 
to illustrate a simple framework for understanding stability in a given DFT-based modeling. We sought the goals to facilitate modeling from hard to soft matter (inside our new XC tool chain). Here we
have 1) Coded the spin-vdW-DF stress result in \textsc{Quantum Espresso} to enable variable-cell vdW-DF calculations in spin systems and 2) used a simple stability condition to discuss soft modes
in \sto, as an example. Having access to a simple, generic, stability gauge 
means that DFT practitioners
have an option of seeking the most relevant
DFT input (controlled by the XC choice) before proceeding with advanced modeling.

We have documented the spin-stress method
contribution for magnetic elements 
as well as for a magnetic perovskite. 
In addition, we have also provided hard- and soft-matter illustrations of using the new tool chain of consistent-vdW-DF XC functionals on benchmarks, a new test of CX performance on DNA intercalation, and a ferroelectric-polymer application of the CX version.

Overall, we find that we will in general need more than just the CX part of the new nonlocal-correlation XC tool chain to cover materials from hard to soft. The AHCX improves the description of the magnetic Fe element
over CX and both CX0P and AHCX are strong performers in our bio-relevant molecular benchmarking. However, hybrid vdW-DFs are not universally improving descriptions either, as discussed for \sto. 

We present these results by looking at a number of 
hard and  soft material cases, in the hope that they
may stimulate further work and analysis.
Studies using different -- but closely related -- regular/hybrid vdW-DFs are interesting not only because they do give useful results and, overall, accurate predictions. The closeness in 
the XC nature of our tool chain means that 
variations in performance may teach us to 
better weigh the balance (and screening)
of truly nonlocal exchange in combination
with our truly-nonlocal-correlation vdW-DF framework. We therefore intend to use of the consistent-vdW-DF tool chain more broadly 
to continue to gather performance statistics.
Ultimately we aim to learn to better identify, \textit{a priori}, the best DFT tool for a given material challenge.

\section*{Acknowledgements}

We thank G{\"o}ran Wahnstr{\"o}m and Gerald D. Mahan for useful discussions.
Work is supported by the Swedish Research 
Council (VR) through  Grants  No.\  2014-4310, 
2016-04162, 2018-03964, and 2020-04997, the Swedish Foundation for Strategic research (SSF) through grants IMF17-0324 and SM17-0020, Sweden's Innovation Agency (Vinnova) through project No.:2020-05179, as well as the Chalmers Area-of-Advance-Production theory activities and the Chalmers Excellence Initiative Nano. The authors also acknowledge computer allocations from the Swedish National Infrastructure for Computing (SNIC), under Contracts  SNIC2019-1-12, SNIC2020-3-13,  and from the Chalmers Centre for Computing, Science and Engineering (C3SE).

\appendix

\section{Coordinate transformations and stress}

We consider small deformations, such that the displacements can be expressed in terms of the strain tensor $\varepsilon_{\alpha,\beta}$. This tensor describes coordinate transformations or scaling
\begin{equation}
\tilde{r}_{\alpha} = \sum_{\beta}(\delta_{\alpha,\beta} + \varepsilon_{\alpha,\beta} )\, r_{\beta} \, .
\end{equation}
Here the subscripts, $\alpha$ or $\beta$, identify cartesian coordinates of the position vector $\bm{r}$. We seek to express 
the resulting stress that arises from the spin formulation $E_{\rm c}^{\rm nl,sp}$
of the nonlocal-correlation energy. 
This stress is formally given as a strain derivative
\begin{equation}
    \sigma_{{\rm c},\alpha,\beta}^{\rm nl,sp}  \equiv 
    - \frac{1}{V}\frac{\delta E_{\rm c}^{\rm nl,sp}}
    {\delta \varepsilon_{\alpha,\beta}} \, ,
\end{equation}
where $V$ denotes the unit-cell volume.

We therefore aim to track every way that the coordinate scaling affects Eq.\ (\ref{eq:cnleval}), for example, through the double spatial integrations, from the spin-density components
$n_{s=\uparrow,\downarrow}(\bm{r})$, and from the
kernel $\Phi$ dependence on coordinate separation $D$. We also need to track the stress
that arises because the local inverse length scale 
$q_0(\bm{r})$ (inside $\Phi$) depends on the spin-density 
gradients $\nabla n_{s=\uparrow,\downarrow}(\bm{r})$. These
gradients change with coordinate scaling because scaling implies
both a density change and a change in taking the derivative with 
positions.
The approach is simply to apply the chain rule for derivatives 
with strain.

The transformation Jacobian is to lowest order
\begin{equation}
    J=\left|\frac{d\tilde{\bm{r}}}{d\bm{r}}\right| = 1+\sum_{\alpha} \varepsilon_{\alpha,\beta}  \, ,
\end{equation}
and corresponds to the strain derivative $\partial J/\partial \varepsilon_{\alpha,\beta} =\delta_{\alpha,\beta}$. Since $E_{\rm c}^{\rm nl}$ involves a double integration, this volume effect
alone produces a the stress contribution 
$ 2 E_{\rm c}^{\rm nl}
\delta_{\alpha,\beta}$ in Eqs.\ 
(\ref{eq:newStress}) and (\ref{eq:Sabatini}).

The kernel $\Phi$ in Eq.\ (\ref{eq:cnleval}) contains a term
that depends explicitly on the separation $D$ between two positions of the electron spin-density distributions. That term resembles the Hartree (or mean-field Coulomb) energy and gives a stress component defined by the second 
row of Eq.\ (\ref{eq:newStress}) and Eq.\ (\ref{eq:SabatiniD}).
Of course, for the spin-polarized case, one must evaluate 
the kernel derivative $\partial \Phi/\partial D$ at inverse length scale values, $q_0(\bm{r})$ and $q_0(\bm{r'})$, 
for the actual spin-density distributions $n_{s=\uparrow,\downarrow}(\bm{r})$. However, that
information is already available from any computations
of the spin vdW-DF description of the nonlocal XC energy
and XC potential.

In reciprocal space, the scaling is given by the transpose of $-\varepsilon_{\alpha,\beta}$. For example, a reciprocal lattice vector scales as
\begin{equation}
    \tilde{G}_{\alpha} = \sum_{\beta} (\delta_{\alpha,\beta} - \varepsilon_{\beta,\alpha})G_{\beta} \, .
\end{equation}
It can readily be shown that with a planewave basis for wavefunctions
$\Psi_{\bm{k},j}= \sum_{\bm{G}} c_{\bm{k}-\bm{G}}^{(j)}
\exp(-i(\bm{k}-\bm{G})\cdot \bm{r})$ there are cancellations
of strain effects in all but the normalization factors
$c_{\bm{k}-\bm{G}}^{(j)}$. The spin-density components
$n_{s=\uparrow,\downarrow}$ will therefore scale with derivatives
given by the volume factor
\begin{equation}
    \frac{\partial n_s(\bm{r})}{\partial \varepsilon_{\alpha,\beta}} = -\delta_{\alpha,\beta} n_s(\bm{r}).
\end{equation}
However, following the logic of the original Nielsen and Martin analysis \cite{Nielsen1985}, the volume scaling of the densities can  be summarized in terms of the relevant (spin-resolved) components $v_{{\rm c},s}^{\rm nl} n_{s}(\bm{r})$ of the XC potential.  The stress terms $- \delta_{\alpha,\beta} \sum_s v_{{\rm c},s}^{\rm nl} n_{s}(\bm{r})$ in Eqs.\ (\ref{eq:newStress}) 
and (\ref{eq:Sabatini}) 
summarize all of the density-volume
scaling effects for the spin-balanced and spin-polarized
cases, respectively.

This finally brings us to the third row of Eq.\ (\ref{eq:newStress}). Here we capture
the effects of strain scaling of the spin density 
gradient $\nabla n_s(\bm{r})$, assuming a fixed 
$n_s(\bm{r})$ variation:
\begin{equation}
    \frac{\partial n_s(\bm{r})}{\partial r_\alpha} \to 
    \frac{\partial n_s(\bm{r})}{\partial \tilde{r}_\alpha}  \approx 
    \frac{\partial n_s(\bm{r})}{\partial r_\alpha} 
    - \sum_{\beta} \varepsilon_{\alpha,\beta}
    \frac{\partial n_s(\bm{r})}{\partial r_\beta}\, .
\end{equation}
For the length of this derivatives we have
\begin{equation}
    \frac{\partial |\nabla n_s(\bm{r})|}{\partial \varepsilon_{\alpha,\beta} }
    = - \frac{1}{|\nabla n_s(\bm{r})|} 
    \frac{\partial n_s}{\partial r_\alpha}
    \,
    \frac{\partial n_s}{\partial r_\beta} \, ,
\end{equation}
because we have handled the volume scaling
of the density separately \cite{Nielsen1985}. The third row of the
spin-vdW-DF stress description Eq.\ (\ref{eq:newStress})
follows by a simple application of the chain rule.

%

\end{document}